\documentclass[final,3p,times]{elsarticle}

\usepackage{amsmath} 
\usepackage[capitalise]{cleveref}
\usepackage{lineno}
\begin{document}

\begin{frontmatter}

\title{The Mechanics and Physics of Twisted and Coiled Polymer Actuators}

\author[inst1]{Qiong Wang}
\author[inst2]{Anan Ghrayeb}
\author[inst3]{SeongHyeon Kim}
\author[inst1]{Liuyang Cheng}
\author[inst1]{Sameh Tawfick}

\affiliation[inst1]{organization={Department of Mechanical Science and Engineering, University of Illinois at Urbana-Champaign },
            city={Urbana},
            postcode={61801}, 
            state={IL}
            }

\affiliation[inst2]{organization={Department of Mechanical Engineering, University of Michigan },
            city={Ann Arbor},
            postcode={48109}, 
            state={MI}
            }

\affiliation[inst3]{organization={Department of Mechanical Engineering, Massachusetts Institute of Technology  },
            city={Cambridge},
            postcode={02139}, 
            state={MA}
            }

\begin{abstract}
Twisted and coiled polymer actuators (TCPAs) generate large contractile mechanical work mimicking natural muscles, which makes them suitable for robotics and health-assistive devices. Understanding the mechanism of nylon TCPA remains challenging due to the interplay between their intricate geometry, chirality, residual stresses, and material microstructure. This study integrates a material microstructure model with rod theory to analytically predict the equilibrium helical shape of the nylon TCPA after fabrication and to explain the observed contraction mechanism upon stimulation. The first ingredient of the model is to treat nylon as a two-phase thermomechanical microstructure system capable of storing strain energy and exchanging it among the two phases. This is validated by characterizing the torsional actuation response of twisted and annealed nylon fibers. The second ingredient of the model is to use the classic Kirchhoff Rod Theory and add a necessary term that couples the bending and twisting energy. Validation with experiments shows that the model captures the equilibrium and longitudinal stiffness of the TCPA in both active and passive states, and the stimulated contraction under external load. Importantly, the model quantifies the influence of the stored energy level on the actuation performance. These concepts can be extended to other types of TCPAs and could enable new material design.

\end{abstract}

\begin{highlights}
\item A new mechanism is proposed for nylon twisted and coiled polymer actuator (TCPA): Twisting stores large strain energy in materials with at least two phases, which can be released as actuation work when stimulated.
\item A mathematical model based on Kirchhoff Rod Theory with a two-phase microstructure captures this new actuation mechanism and is validated by experiments.
\item The proposed mechanism is different than the existing explanations which are based on swelling of twisted and coiled rods. The new mechanism considers the effect of stored elastic energy and its release on actuation.
\item The proposed model can be extended to search for new polymers offering superior actuation performance.
\end{highlights}


\end{frontmatter}

\section{Introduction}
\label{sec: Intro}
Twisted and coiled polymer actuators (TCPAs) are elastic actuators that mimic the structure of biological muscles via helical twisted and coiled synthetic fibers \cite{Haines2014ArtificialThread, Lee2017ElectrochemicallyMuscles}. Due to their fiber-like form and longitudinal contraction when stimulated, they are known as coiled artificial muscles. They have become increasingly popular due to their ability to generate large forces and strokes while being lightweight, low-cost, scalable, and flexible. TCPAs can be manufactured from a variety of materials, ranging from polymers to carbon fiber composites, while the manufacturing method is relatively simple and universal \cite{Haines2014ArtificialThread,Lima2015EfficientYarns,Chen2015HierarchicallyVapours,Kim2016Bio-inspiredMuscles,Yang2016ArtificialFibers,Lee2017ElectrochemicallyMuscles,Yang2018,Jiang2021}.

Twisting and coiling provide a great benefit in transforming intrinsic stimuli responsiveness on the material level to longitudinal contraction, which is the primary deformation observed in nature. The precursor fiber is twisted first and then coiled into a helical shape. The helical shape is unstable after room-temperature coiling and it would untwist considerably if the tether is removed. However, the shape can be fixed by thermal annealing in the case of nylon, or by coating with an external layer in the case of carbon nanotube fibers. Extensive research focused on developing TCPAs with improved performance and their applications \cite{Cui2023Pretension-FreeCapability,Hu2023BioinspiredActuators,  bao_fast_2023,wu_construction_2024,li_effective_2024,shin_woven_2024,matharu_jelly-z_2023,matharu_single_2024,hu_artificial_2024,weerakkody_modeling_2023}. Active materials such as shape memory polymers and liquid crystal elastomers are used to produce new powerful actuators based on twisting and coiling \cite{Gao2023LiquidMuscles,Cui2023Pretension-FreeCapability,Lugger2023Melt-ExtrudedActuators,zhang_compound_2024,noh_high_2024,escobar_fast_2024}. This variety of materials serves as a strong motivation towards understanding the underlying mechanism governing the actuation mechanisms, which could enable new material designs on the polymer microstructure level. 

Several models have been proposed to explain the mechanism of TCPAs based on swelling-induced untwist. This mechanism can be considered in two cases: tethered and untethered ends. The first and simplest model from Haines \textit{et al.} \cite{Haines2014ArtificialThread, HainesNewTwist} proposed a kinematics model for tethered actuation based only on geometry. To provide qualitative insight without considering the complex mechanics of materials, the authors assumed that the fiber undergoes radial swelling in response to external stimuli. This swelling is then transformed into fiber untwist due to kinematic constraints. They also assumed that the elastic moduli do not change during heating. According to the classic Kirchhoff Rod theory, the total twist of the helix can be derived from the helix angle and the coil diameter. Thus with the amount of untwist, one can estimate the kinematic contraction. This simple model qualitatively explains the actuation provided by the TCPA. However, it is based on the kinematic assumption that the change in internal twist is equal to the change in geometry torsion \cite{LOVE1892}, which is an unrealistic assumption if the elasticity is considered. 

A study by Sharafi \textit{et al} \cite{Sharafi2015AMuscles} considered the micro-structural features of the semi-crystalline material and proposed a phenomenological thermomechanical constitutive model consisting of helical chains, entropic chains, and thermo switches. This model provides a quantitative prediction of the intrinsic actuation of the material based on the microstructure descriptions. However, this prediction requires either molecular scale parameters, which are challenging to measure, or the use of many fitting parameters. In particular, the geometric effect of a twisted helix is not considered in this work. The contribution of crystalline microphase is also assumed to be negligible.

Combining both geometry and microstructural aspects, a more elaborate multiscale model was developed by Yang \textit{et al} \cite{Yang2016b}. The untwist and recovered torque is provided by the concentric multilayer anisotropic laminate, whose bias angle changes along the radial direction. The stroke of the TCPA was then determined for the spring using Castigliano's second theorem. Furthermore, the finite element analysis (FEA) model extends Yang's approach to a numerical framework, offering a more adaptable and versatile tool to predict the performance of TCPAs \cite{Hunt2021Thermo-mechanicalFEA,Liu2024MechanicsSimulation}. These two models can provide quantitative prediction to the actuation performance of TCPAs with limited fitting needed and good agreement with the experimental results.

While the models for tethered-end actuation have been extensively studied, untethered actuation is less explored. With tethered-end, radial swelling generates a torque due to fiber untwist, and this recovered torque is integrated into Kirchhoff rod theory and Castigliano's theorem to determine the contractile actuation. This mechanism is especially relevant when explaining the actuation of TCPAs which indeed require tethering such as the case of CNTs or carbon fibers embedded in a swelling matrix or the case of an anisotropic swelling material due to its aligned molecular microstructure. Hence, by definition, to model this "recovered torque", the boundary conditions must prevent untwist. This precludes the applicability of these models for explaining the untethered actuation of nylon or core-shell TCPAs such as the sheath-run artificial muscles and has been investigated in other studies \cite{leng_tethering_2023}. For the second case where the coil is untethered, analytical models become quite challenging if not impossible to apply. Finite element analysis (FEA) can be used to model swelling-based mechanisms in untethered nylon TCPA. However, this method has some disadvantages that are described in the introduction. For example, the experimentally observed softening of TCPA and the effect of elastic energy storage are not intrinsically included in the mechanism but must be added through additional assumptions.

Apart from boundary conditions, there is one extremely important aspect that has not been captured in any existing model. That is the critical role of the strain energy stored due to the internal twist of the rod. During the manufacturing process, the fiber is twisted up to 1200 rad/m, which is equivalent to 60 \% strain at1 its surface for a 1 mm diameter cylindrical fiber. For the existing models, the effect of twist has been only considered as a geometry constraint of the concentric multi-layer laminate. It has been assumed that all the elastic energy stored in the fiber during the twisting process is completely dissipated during the annealing process. However, this is not guaranteed, since the typical annealing temperature of nylon TCPA is lower than its melting point and the deformation of the crystalline part might not be fully released. The role of internal residual strain energy leads to a new hypothesis about the actuation mechanism of nylon TCPA. The hypothesis, which guides this study, is discussed next. 
 
 Given the stated constraints and assumptions, it appears that the swelling-based mechanism falls short of providing a comprehensive explanation for the actuation of TCPA. However, it is important to note that this explanation may not be the sole perspective. Another concept suggests that the actuation of TCPA may be linked to the stored elastic energy and released strain within the twisted helical structure \cite{Yuan2014,Whitesides2018}. While the modeling of TCPA has only gained attention recently, the modeling of twisted helical rods has interested mathematicians for over 150 years. In the 1860s, Gustav Kirchhoff formulated the renowned Kirchhoff's kinetic analogue \cite{G.Kirchhoff1859UeberStabes.}, demonstrating that a thin, prismatic, and straight rod subject to terminal forces and torques can be analogously represented as a rigid body rotating about a fixed point. Cosserat rod, a generalization of Kirchhoff rod theory is now commonly used in the simulation of complex rod structures. It comes to us that this system captures the essence of the actuation of the Nylon TCPA. It can describe the geometry change during the loading and actuation process. It also includes the effect of the stored elastic strain energy which is not considered in the swelling-based mechanism. However, current study on the twisted helix focuses mostly on investigating the instability of the helical structure \cite{Thompson1996,VanDerHeijden2000,Chouaieb2006Helices,CrossedDSignurickovic2013TwistFilaments,Borum2020WhenStable,Gomez2023TwistingTransition,Liu2024MechanicsSimulation}, yet there is a notable gap in understanding the behavior of this system in parametric space, such as temperature.

In this article, we study a new hypothesis for the actuation of coiled artificial muscles: The coiled artificial muscle stores elastic energy and is a system under equilibrium - it exerts mechanical work during actuation since the change of the external condition shifts the equilibrium of the system and leads to a higher work potential. We develop an analytical model that captures the important effects of finite geometric changes, residual stresses, and material microstructure. The key ingredient of this study is to treat the TCPA as an elastic rod system with stored elastic energy under equilibrium whose parameters are subjected to environmental changes. During the manufacturing process, we store a large amount of strain energy within the structure and fix the system at that state. Upon heating, the system parameters change and thus shift the equilibrium configuration, providing the stroke. To formulate this system analytically, we introduce a hybrid model that integrates a microstructure model for highly strained semicrystalline materials with a geometrically nonlinear model for elastic rods. The foundation of this model lies in the well-established Kirchhoff Rod Theory and the principle of minimum total potential energy (MTPE) which describes the equilibrium state of the twisted helix. Then, we incorporate a microstructure model inspired by shape memory polymers and Takayanagi Model for semi-crystalline materials \cite{Takayanagi2007}. To validate the proposed model, thermo-torsional tests are conducted on straight, twisted, and annealed fibers. Finally, the coupling between the bending and torsional modes is considered to account for the off-axis polymer chains generated due to twisting and annealing. The added term ensures that the system is solvable under any combination of external force and torque. With the complete model, the actuation can be provided even with the absence of the end-tethering. The accuracy of the model is finally assessed by comparing the results with DMA tests and isobaric water tests on the TCPA, which reveals satisfactory agreement. The force-displacement curve results are also validated, showing that the model also works in the non-actuated state of the TCPA. In contrast to the swelling-based model, the model introduced in this study offers a two-phase microstructure model that uses the intrinsic polymer physics behavior to capture the helical stability of untethered annealed TCPAs and their observed macroscopic softening. It provides an alternative actuation mechanism involving the 'untwisting' and shifting of equilibrium through changes in the energy-storing microstructure, rather than relying on geometric constraints. This model adds up to existing mechanisms by enabling free-end actuation and expanding material choices for better TCPAs, transitioning focus from anisotropic thermal expansion materials to extensively researched shape memory polymers and liquid crystal elastomers.

The outline of this paper is as follows. In \cref{sec: fabrication}, we introduce how the TCPA samples are prepared for the tests in this work. In \cref{sec: Thermo-Mech Test}, we present the experimental tensile test response with actuation performance curves and explain the unusual response observed in the tests. Then \cref{sec: Love Model} investigates how to model the complex geometry of TCPA based on Kirchhoff Rod Theory and minimal potential energy method. \cref{sec: Role of Micro} investigate how to model the material properties of the highly strained semi-crystalline polymer. Finally in \cref{sec: Complete Math Model}, we combine the models that have been proposed in previous sections with the coupling effect to construct a complete model for the TCPAs. Model validation is presented in \cref{sec: Model validation} where two different tests show good agreement with the simulation. \cref{sec: Discussion} provides discussions on the distinctive features of the proposed model, its limitations, and resulting insights.

\section{TCPA Fabrication}
\label{sec: fabrication}
In this section, we will introduce the fabrication method of the TCPA specimen. We first twist and coil the monofilament nylon fiber to form a twisted helix. The fiber diameter and the geometry of the helix can vary depending on the need for power and stroke (\cref{fig: DMATest}a). In this work, all of the samples tested are made from fibers with 1 mm diameter and 400 mm length. Fiber (from \textit{Shaddock Fishing}, 1 mm clear monofilament fishing line, 57.47 kg test strength) has a density of 1 g/cm$^3$. The fiber is first twisted by 88 turns with a 1 kg load applied at one end to prevent unwanted local writhing. Note that the amount of deformation is above the limit of the elastic region, and the fiber undergoes plastic deformation. The fiber is only able to recover around a quarter of the initial deformation if released. This plasticity is not quantitatively characterized in this model since its effect mixes up with the effect of annealing and is hard to distinguish. In the mathematical model, they will be compensated by a constant shift to the strain field. Then, the twisted fiber is wrapped around a metal pin with a diameter of 2.5 mm to form a helix. The shape of the twisted and coiled fiber is fixed by annealing it in an oil bath at 170 ° C for 7 hours. 

\section{Thermomechanical Testing}
\label{sec: Thermo-Mech Test}
In this section, we will perform a series of thermomechanical tests on TCPAs to characterize their tensile response and the actuation behavior.

TCPAs, when activated, undergo property changes in response to environmental conditions. This inherent characteristic is a common feature of soft actuators, as they dynamically interact with their surroundings, adapting their response accordingly. Therefore, it is important to start the evaluation of the actuator with a mechanical test. Since TCPA is activated by heat, a thermomechanical analysis is needed.  The results reveal that the actuator provides actuation while undergoing softening. This observation is not unexpected considering that nylon softens upon heating, but it is intriguing, as a softer structure demonstrates the capacity to handle loads and provide mechanical work.

\begin{figure}[t]
\centering
\includegraphics[width=1\textwidth]{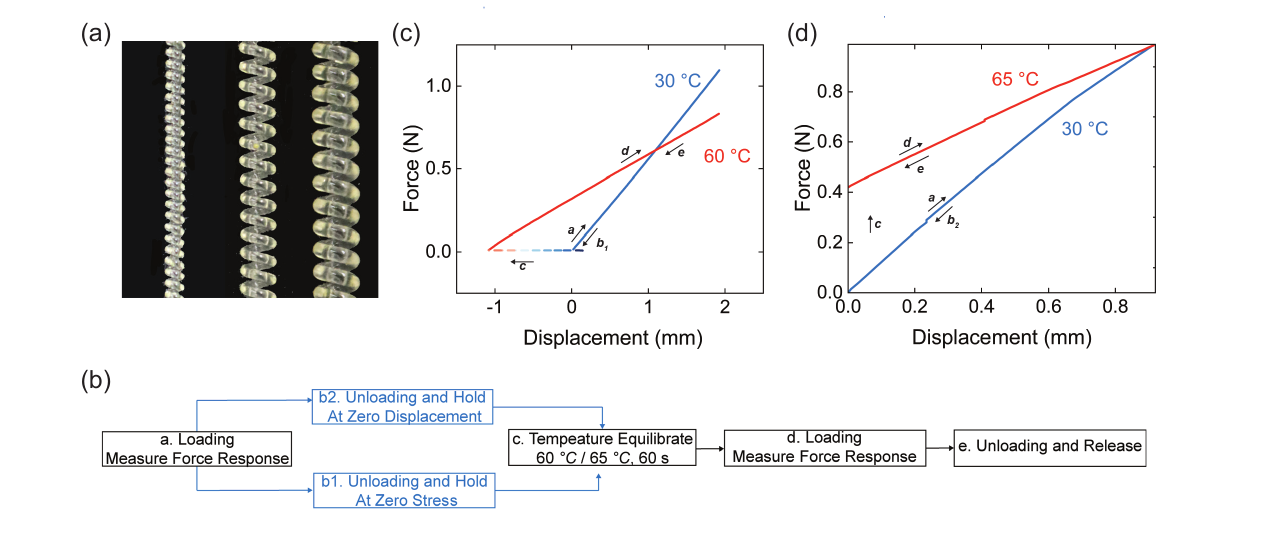}

\caption{Structure and thermomechanical analysis of TCPA;
(a) Optical images of TCPA from nylon fishing lines with different spring indices;
(b) Test procedure of the DMA test. The furnace is first equilibrated at 30 $^\circ C$ and the initial length of the sample is determined with a preload of 0.01 $N$. The tensile test (step a) is then conducted. Following the first test, two different procedures are conducted. The first is to unload the sample to its initial length (step b2), after which it is maintained at that position. The second is to unload the sample until 0.01 $N$ (step b1), after which the force remains at that small preload value throughout the heating process. Subsequently, the furnace is heated up to 60 $^\circ C$ or 65 $^\circ C$ and equilibrated for 60 s (step c). The second tensile test (step d) is then performed;
(c) The force-displacement curve for a TCPA at lower temperature (blue) and actuation temperature (red) measured as described in (b). The heating process is performed with zero external load; d)The force-displacement curve for a TCPA at a lower temperature (blue) and actuation temperature (red) measured as described in (b). The heating process is performed with zero displacement load.  The unloading data is not plotted.}
        \label{fig: DMATest}
\end{figure}

To characterize the behavior of the TCPA at different temperatures, thermomechanical tests of the TCPA sample are conducted by a dynamic mechanical analyzer (DMA 850, TA Instrument) with the standard tension clamp. \cref{fig: DMATest}b illustrates the experimental procedure of the uniaxial tensile test. The test results (\cref{fig: DMATest}c-d) indicate simultaneous contraction and softening of TCPA upon heating. The response of the sample can be approximated by linear curves with distinct slopes and y-intercepts. Both results show similar trends. The slope at higher temperature indicates that the stiffness of the sample reduces to half its value compared to the case at the lower temperature. Additionally, the y-intercept signifies that the sample contracts upon heating, resulting in a response force or stroke depending on the loading condition.

The results lead to two observations concerning the mechanism of the TCPAs. First, the actuation is primarily induced by a shift in the rest length caused by the temperature change. The \textit{c} curve in \cref{fig: DMATest}c shows the length of the sample upon heating without axial constraints. It indicates that the actuator at 60 $^\circ C$ has a shorter rest length compared to the actuator at 30 $^\circ C$. Similar thing is observed in the test shown in \cref{fig: DMATest}d. The reaction force is generated as shown in curve \textit{c} since the end of the sample is held at zero displacements and prevented from actual contraction. Secondly, the softening of the actuator is not negligible. This phenomenon is counter-intuitive since one would naturally expect the actuator to be stiffer or, at the very least, retain its initial stiffness when it lifts a load. Existing theories often explain the mechanism of TCPA by focusing on the untwist resulting from the radial swelling of the fiber while the softening effect is less discussed. They tend to overlook the change in elastic modulus so as to simplify the analysis. However, the softening effect is technologically important as it influences the actuation performance of the actuators when attached to an elastic structure. Notably, Chen et al.\cite{chen_effect_2023} have conducted a thorough investigation into how temperature affects the recovered torque of the swelling-based mechanism. In the following section, we will explore the impact of softening on the geometry of a simple helical equilibrium based on the rod theory which sets the foundation of the proposed mechanism in this work.

\section{The Role of Twisted Helical Geometry}
\label{sec: Love Model}
 In this section, we will study the TCPA structure based mostly on the classic Kirchhoff-Love rod theory \cite{LOVE1892} and its variation form for the non-isotropic rod \cite{Chouaieb2006Helices}. We will start the analysis by revisiting the classic helical solutions of the terminal-loaded rod with an emphasis on the counterintuitive behavior of a helix caused by the twist. Following that, we will demonstrate the behavior of the twisted helical rod as the temperature varies and how that could lead to powerful actuation.

\subsection{Quadratic Strain Energy and Equilibrium}
\label{subsec: Strain Energy}
\begin{figure}[t]
\centering
\includegraphics[width=1\textwidth]{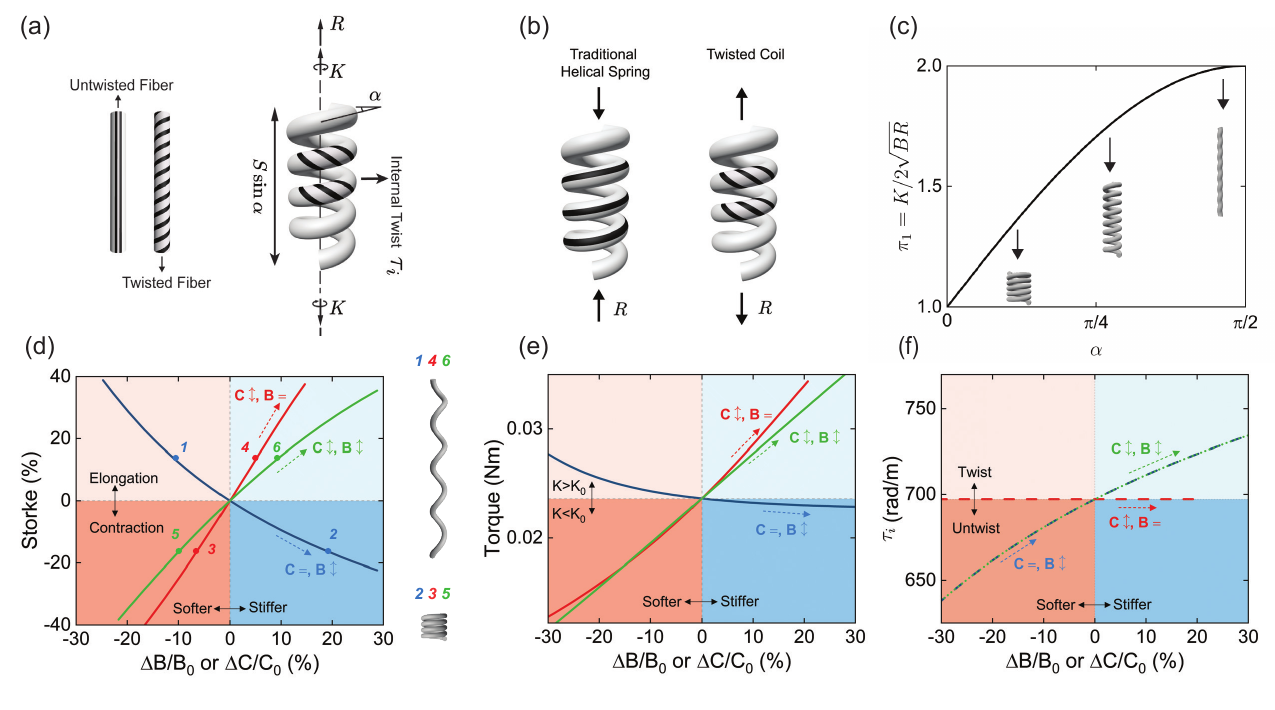}

        \caption{TCPA is modeled as twist and stretched helical rod under Kirchhoff-Love theory;
(a) Schematic diagram of a straight fiber, twisted fiber and TCPA (twisted helix);
(b) Force needed to deform a straight rod to a traditional helical spring or a twisted coil. The direction of the force in theses two cases are opposite to each other. If the bending mode dominates, a compression force is needed to hold the shape. If the twisting mode is dominated (twisted coil), a stretching force is needed to hold the shape;
(c) Plot of non-dimensional parameter $\pi_1$ versus helix angle $\alpha$ with the shape of the correlated helix shown in the inset figures;
(d-f)Stroke, torque and internal twist of the helical rod under different bending and torsional stiffness. The x-axis is the change of moduli in percentage and the y-axis is the stroke, reaction torque, and the internal twist measured from the twisted helix. The orange region indicates a decrement in rigidities which commonly happens during heating. The blue region indicates an increment in rigidities which happens more during cooling.
}
        \label{fig: TCPASpring}
\end{figure}

In this subsection, we will describe the strain field and the strain energy of the twisted helix from its geometry parameters. Then, we show how the equilibrium configuration of the helix can be calculated by taking the variation of the total potential energy.  

Consider a thin straight rod with length $S$ bent and twisted into a helical shape with $N$ coils and a total helix length of $L$ (\cref{fig: TCPASpring}a). The details of TCPA's geometry can be found in the Appendix A. Its helix angle and internal twist angle are denoted as $\alpha$ and $f$, and its unit fiber length is defined as $l=\frac{L}{2\pi N}$. The centerline of the rod can be described by 
\begin{align}
    \mathbf{r}(s)=\{\frac{l}{\cos \alpha} \cos l s,\frac{l}{\cos \alpha}  \sin l s, \frac{l}{\sin \alpha} l s\} \label{eqn: ArcPara},
\end{align}
where $s$ is the arc length.

Assuming a point traveling along the central axis of the rod with unit velocity, the principal torsion-flexure axes corresponding to that point will also rotate. The angular velocities along the axes with which they rotate are the components of strain $\mathbf{u}$ - curvature $\kappa_1,\kappa_2$ and total twist $\tau$ at that point \cite{LOVE1892}:
\begin{align}
&\mathbf{u}=\{\kappa_1,\kappa_2,\tau\} ,\kappa_1=\sin f\ \kappa,\kappa_2= \cos f\ \kappa,  \tau= \tau_i+\tau_s\label{eqn: Tau},\\
& \kappa =\frac{\cos\alpha }{l} ,\tau_s=\frac{ \ \sin \alpha}{l} .
\end{align}
Here we adopt the nomenclature from Thompson \cite{Thompson1996} that the total twist $\tau$ is composed of two parts. The first part is the internal twist $\tau_i$. Consider a straight rod twisted by angle $f(s)$ along the axial direction, the twist applied to the local infinitesimal rod can be defined as the derivative of the internal twist angle $f(s)$, which is $\partial f$/$\partial s$. In the case of TCPA discussed in this work, the twist angle can be approximated as a linear function of the internal twist $f=\tau_i s$, giving a constant value of $\tau_i$. The second part is the geometric torsion $\tau_s$, a measure of the tortuosity of the central line, similar to the torsion defined in a Frenet-Serret formula. These two components are interchangeable under deformation. For a normal engineering spring without internal twist, the total twist is directly given by the geometric torsion $\tau=\tau_s$. As the spring is stretched, it is straightened and the total twist is solely contributed by the internal twist $\tau=\tau_i$ because a straight rod is not tortuous.

The total potential energy of a conserved elastic system for a given set of kinematically admissible displacement field $\mathbf{u}$ is composed of the elastic strain energy $U$ stored in the system and the work potential $W$. The minimum potential energy method states that although there may be infinitely many solutions that satisfy a particular boundary condition, only equilibrium points can be observed physically:
    \begin{align}
    \delta{\Pi}=\delta U+\delta W=0 \quad \text{for all admissible variations } \delta\mathbf{u}. \label{eqn: EquilEqnPartial}
\end{align}
The structure of a twisted helix can be completely described by three independent parameters: helix angle $\alpha$, internal twist $\tau_i$, and normalized fiber length per coil $l$. Then the requirement given in \cref{eqn: EquilEqnPartial} becomes that the partial differential of the total potential energy of the three parameters shall be zero to ensure equilibrium:
\begin{align}
    \delta \Pi =0 \to \frac{\partial\Pi}{\partial \alpha}=\frac{\partial\Pi}{\partial l }=\frac{\partial\Pi}{\partial \tau_i}=0 .\label{eqn: PartialCond}
\end{align}

  As indicated in \cref{fig: TCPASpring}a, the work potential of the external load is given by the summation of the work done by the external force $W_R$ and the work done by the external torque $W_K$:
\begin{align} 
    W=W_K&+W_R=-S K(\tau_i+1/l-\tau_{i0}-1/l_0)+SR (\sin{\alpha_0}-\sin{\alpha}), \label{eqn: WorkPotential}
\end{align}
where $\alpha_0$, $\tau_{i0}$ and $l_0$ are the initial conditions. 

Considering that the nylon fiber used by TCPA is a nonisotropic material, we adapt the formulation of the strain energy from Chouaieb et al. \cite{Chouaieb2006Helices}. For an inextensible, unshearable but non-isotropic rod, the strain energy is given as
\begin{gather}
\begin{split}
    U(\mathbf{u})=\frac{1}{2} S (\mathbf{u}-\mathbf{u}_0)\cdot \mathbb{C}(\mathbf{u}-\mathbf{u}_0),\\
    \mathbb{C}=\begin{pmatrix}
B_1 & 0 & C_{13}\\
0 & B_2 & C_{23}\\
C_{13} & C_{23} & C
\end{pmatrix},
\end{split} \label{eqn: StrainEnergy}
\end{gather}
where $S$ is the length of the rod, $\mathbf{u}$ is the displacement field from \cref{eqn: Tau}, $\mathbf{u}_0$ is the strain in the unstressed reference configuration. $\mathbb{C}$ is the symmetric positive-definite elasticity tensor where $B_1$ and $B_2$ are the bending rigidity along the normal and binormal directions, $C$ is the torsional rigidity along the tangent direction, $C_{13}$ and $C_{23}$ are the couplings between bend and twist. 

\subsection{The Counterintuitive Behavior of the Twisted Helix}
\label{subsec: Interesting behavior}

Assuming the classical Kirchhoff rod model, we are able to resolve the system and determine the equilibrium states of a twisted helix. In this subsection, we will explain the behavior of the twisted helix in contrast to a conventional engineering helical spring, and offer an analysis for its role in the actuation. 

After the manufacturing process of TCPA, the straight precursor fiber is twisted and coiled into a twisted helical rod held by terminal force $R$ and terminal torque $K$ (\cref{fig: TCPASpring}a) which can be mathematically described by \crefrange{eqn: PartialCond}{eqn: StrainEnergy}. This twisted helical structure has a qualitative difference compared to the widely used engineering helical spring - the internal twist along the fiber direction. This difference leads to the need for different directions of the force to hold the helical configuration (\cref{fig: TCPASpring}b). In the more intuitive case of transforming a straight rod into a loosely coiled spring, one must bend it into circular shapes that overlap each other with an external force pointing inward. However, when a straight rod is first twisted and then coiled into a twisted helix, the force required to sustain the deformation points outward like an 'inverse' spring. In both cases shown in \cref{fig: TCPASpring}b, when the force is released, the rod will restore its straight geometry. Despite its limited recognition, the explanation for this phenomenon is provided by the Kirchoff-Love rod theory. Here we will prove it by solving the exact solution of the twisted helix. 

For ease of discussion, in this section we consider the isotropic rod which gives $B_1-B_2=C_{13}=C_{23}=0$, leaving matrix $ \mathbb{C}$ a diagonal matrix. An assumption behind this formulation is that there is no coupling between bend and twist which we will discuss in detail in \cref{sec: Complete Math Model}. It is also assumed that the bending has no preferred direction radially. Then the strain energy density from \cref{eqn: StrainEnergy} can be simplified as
\begin{align}
    U=U_B&+U_T =\frac{1}{2} S B(\kappa-\kappa_0)^2+ \frac{1}{2} S C(\tau-\tau_{0})^2, \label{eqn: ElasEnergyLong}
\end{align}
where $\kappa_0=\frac{\cos{\alpha_0}}{l_0}$, $\tau_0=\frac{\sin{\alpha_0}}{l_0}+\tau_{i0}$.

Substituting \cref{eqn: WorkPotential} and \cref{eqn: ElasEnergyLong} into \cref{eqn: PartialCond}, one can simplify and obtain three equilibrium equations:
\begin{align}
\begin{cases}
    R l \cos \alpha=-G \sin \alpha+H \cos \alpha  \\
    K=G \cos \alpha +H \sin \alpha \\
    K=H
\end{cases} ,\label{eqn: ThompsonSolution}
\end{align}
where $G=B\kappa$ is the flexural couple and $H=C\tau$ is the torsional couple. The first two equations satisfy the force and torque balance condition. The third equation is the additional condition that accommodates the requirement of the minimal potential energy. Thompson has pointed out in his study \cite{Thompson1996} that this system has a nonlinear parameter relating the loading conditions and the geometry:
\begin{align}
    \pi_1=\frac{K}{2\sqrt{BR}}=\frac{1+\sin{\alpha}}{2}. \label{eqn: PIExp}
\end{align}
 For a given pair of $K$ and $R$, the corresponding helix angle $\alpha$ can be determined through \cref{eqn: PIExp}, and its result is represented graphically in \cref{fig: TCPASpring}c. The allowable range of $\alpha$ spans from $0$ to $\frac{\pi}{2}$ which is a physically admissible value. The reader is advised to look at \cref{fig: TCPASpring}a simultaneously with \cref{fig: TCPASpring}c to appreciate that only a prescribed combination of force and torque leads to a solution of this nonlinear mechanism. As indicated in \cref{fig: TCPASpring}c, an increase in tension force results in a decreased helix length which leads to the 'inverse' spring behavior aforementioned. This behavior originated from the contribution of the torsional couple $H$. As indicated in \cref{eqn: ThompsonSolution}, $H$ has a positive contribution to the external force $R$ while the flexural couple $G$ has a negative contribution. For a normal engineering spring, curvature $\kappa$ is much larger than torsion $\tau$, resulting in negative $R$ pointing inward as we discussed in \cref{fig: TCPASpring}b. For a twisted helical rod, the internal twist $\tau_i$ is much larger than $\kappa$. Thus, the torsional couple $H$ is much larger than the flexural couple $G$, leading to a positive external force $R$ pointed outward.

Although this is not new information, it is presented here to emphasize the counterintuitive behavior of the twisted helix. In most discussions of the twisted helix, this behavior is often diluted in stability and topology analysis \cite{Borum2020WhenStable, Chouaieb2006Helices,Chouaieb2004KirchhoffsRods,Charles2019TopologyFibers}. The reason we focus on this unusual force response of the twisted helix is that it leads to the potential to operate TCPAs by simply changing the moduli. To explore this possibility, we can investigate the parametric space of the the twisted helix. Guided by the experimental results in \cref{fig: DMATest}c, we explored the effects of the moduli $B$ and $C$. By keeping the boundary conditions the same, one can study how the change of moduli in \cref{eqn: ThompsonSolution} could influence the equilibrium. The stroke provided by the twisted helix can be evaluated by calculating the change of the equilibrium length as the moduli change. We study a twisted helix with an initial helix angle $\alpha_0$, bending rigidity $B_0$, and torsional rigidity $C_0$. To mimic the actual working conditions of the TCPA, the boundary conditions of the test are assumed to be a fixed external load with tethered end. This test is commonly used in the evaluation of TCPAs and is referred to as a fixed-end isobaric test. 

Three different cases have been investigated with results plotted in \cref{fig: TCPASpring}d-f. The first case is to keep the torsional rigidity $C$ constant ($C=$), and change only the bending rigidity ($B\updownarrow$). As the bending rigidity increases, the twisted helix becomes shorter and provides a contractile stroke. This is similar to the behavior of a normal helical spring. With a larger stiffness, the spring can contract and lift an external load. During this process, the external torque provided by the actuator decreases while the internal twist increases. The second case is to keep the bending rigidity constant ($B=$) and change only the torsional rigidity ($C\updownarrow$). In this case, the trend of the stroke is flipped. As the torsional rigidity decreases (softening), the twisted helix will provide a contractile stroke. The torque, on the other hand, increases as the torsional stiffness increases as expected. However, the internal twist is ept constant throughout the process. Similar behavior is observed in the third case where the bending and torsional rigidity change at exactly the same rate ($B\updownarrow, C\updownarrow$), which is the case of a homogeneous material. This is to say that under these cases, while the material is 'softer' in a certain manner either due to heating or other triggers, the twisted helix will lift a constant external load and provide a contractile stroke. As indicated in \cref{fig: TCPASpring}f, the actuation does not solely rely on the change of internal twist, but the change of the equilibrium configuration. By changing the parameters of the system, the available equilibrium is effectively shifted and induces either contraction or elongation upon heating.
 
In summary, the twisted helix having internally stored elastic energy generates a large actuation stroke from a change in rigidity. Specifically, a decrease in torsional rigidity leads to contractile actuation of a twisted helix. This guides the hypothesis of this study by motivating the measurement of the decrease in torsional rigidity upon heating, and the mathematical modeling of the actuation mechanism based on stored elastic energy and residual stresses. This kind of behavior arises due to the fact that the twisted helix is the equilibrium state of a complex nonlinear mechanism (akin to a nonlinear linkage) balanced by the external force and torque. This mechanism is specifically useful in the design of actuators since making material softer is readily realizable by heating or swelling it with a solvent. However, this mechanism has several limitations. The first limitation lies in the boundary conditions. After the annealing process, the twisted helical sample does not need terminal force and torque to hold its shape which is not covered by the current mechanism. Secondly, the theory predicts an 'inverse spring' behavior, as we discussed, while the TCPA behaves more like a normal spring at room temperature. Finally, the model reduces the cylindrical rod into a curved center line. This prevents it from capturing the local fiber 'untwist' due to radial swelling as part of the kinematics of the actuation. With all these limitations, the current model is not ready to quantitatively capture the actuation performance of the TCPA. By definition, since the current mechanism assumes a single-component material, it cannot be equilibrated in a helical shape without a terminal wrench. In the experiments, this new equilibrium at the end of the fabrication and annealing can be explained by the residual stress due to the annealing process. In the next section, we will model it by considering a composite material where the two components are strained to different degrees and show how this enables an equilibrium without the need for external forces or torques.

\section{The Role of Material Microstructure}
\label{sec: Role of Micro}

TCPAs can be made from various materials such as hydrogel \cite{Zhang2021a, Sim2020, Cui2021}, carbon nanotubes \cite{Mu2019Sheath-runMuscles, Song2018HierarchicalMuscles,Lee2017ElectrochemicallyMuscles,Lima2012ElectricallyMuscles,Chen2015HierarchicallyVapours,Lee2014}, niobium \cite{Mirvakili2013NiobiumMuscles}, graphene oxide \cite{Kim2018b,Hyeon2019}, carbon fibers \cite{Lamuta2018TheoryMuscles}, and high-strength semi-crystalline or composite polymer fibers \cite{Haines2014ArtificialThread,Ravandi2014SimpleHeating,Yang2016ArtificialFibers,Aziz2021,Sim2020,Huang2020}. Among them, high-strength semi-crystalline fibers such as commercial monofilament fishing lines made from nylon possess advantages in manufacturing cost, production stability, wide operating range, and high work capacity. In this work, we focus on the TCPAs made from nylon fibers. Although the twisted and coiled geometry is the essential component of TCPAs, their material properties are as important to ensure the actuation. With the model introduced in \cref{sec: Love Model}, we are able to explain why the twisted and coiled structure can provide the actuation under heating. However, the model is not able to capture the process of annealing since the Kirchhoff rod theory is not designed for multi-phase materials such as semi-crystalline polymers or composite materials. With the annealing procedure, the shape is fixed as the amorphous phase passes through the glass transition temperature. Thus, an appropriate model to count for this effect is necessary to predict the performance of nylon TCPAs completely. 

To investigate how the microstructure and material properties influence actuators, thermomechanical tests are performed. Since complex helical geometry will add unnecessary complication to the analysis, we will perform all the following tests on straight fibers. Two different torsional tests are designed to determine the torsional thermomechanical behaviors of the sample under heating and twisting conditions. Based on the test results, we will first discuss how the twisted fiber and TCPAs share similarities with two-way shape memory polymers. Finally, a mathematical microstructure model is proposed. The material properties of the proposed model are determined by the thermomechanical measurements.   

\subsection{Thermomechanical Measurement of Twisted Fibers (Non-coiled)}\label{subsec: Thermo-Mech Twisted}

\begin{figure}[t]
\centering
\includegraphics[width=1\textwidth]{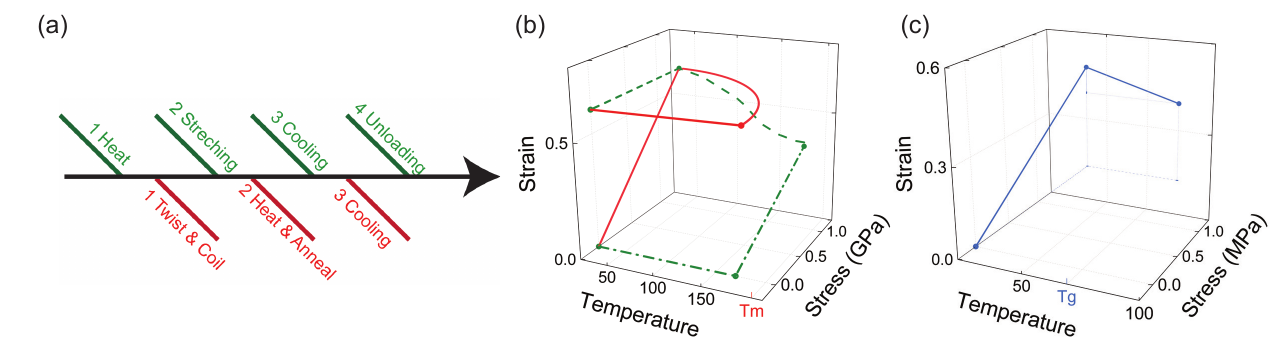}

        \caption{ Manufacturing (Programming) and actuation process of TCPA and shape memory polymer. The numerical values adopted in the graph are based on the typical TCPA manufacturing/actuation process and can differ for different two-way shape memory polymers; 
        (a)Manufacturing (programming) steps of TCPA (red) and shape memory polymer (green); 
        (b) Strain-temperature-stress curve of TCPA (red solid line) and shape memory polymer (green dashed line) during manufacturing (programming); 
        (c) Strain-temperature-stress curve of TCPA and shape memory polymer during actuation. Step 1 is room temperature stretching and step 2 is actuation with temperature change. Hysteresis is not plotted in this graph.  }
        \label{fig: ProgrammingSMPTCPA}
\end{figure}
In this sub-section, thermomechanical torsion testing is conducted on the twisted and annealed fiber. Looking at \cref{eqn: ThompsonSolution} and \cref{fig: TCPASpring}, the torsional rigidity plays a major role in the actuation. Since the nylon monofilament fiber is first twisted and then annealed, its torsional rigidity can differ significantly from the case of the precursor fiber. One clear evidence is that the fiber will reversely rotate when it is heated up and cooled down. Two thermomechanical tests are then designed to study the aforementioned properties. The first test is the fixed-end reaction torque test which measures the mechanical response of the sample under angular displacement at varying temperatures. The second test measures the free-end rotational response of the sample under varying temperatures.

\begin{figure}[t]
\centering
\includegraphics[width=1\textwidth]{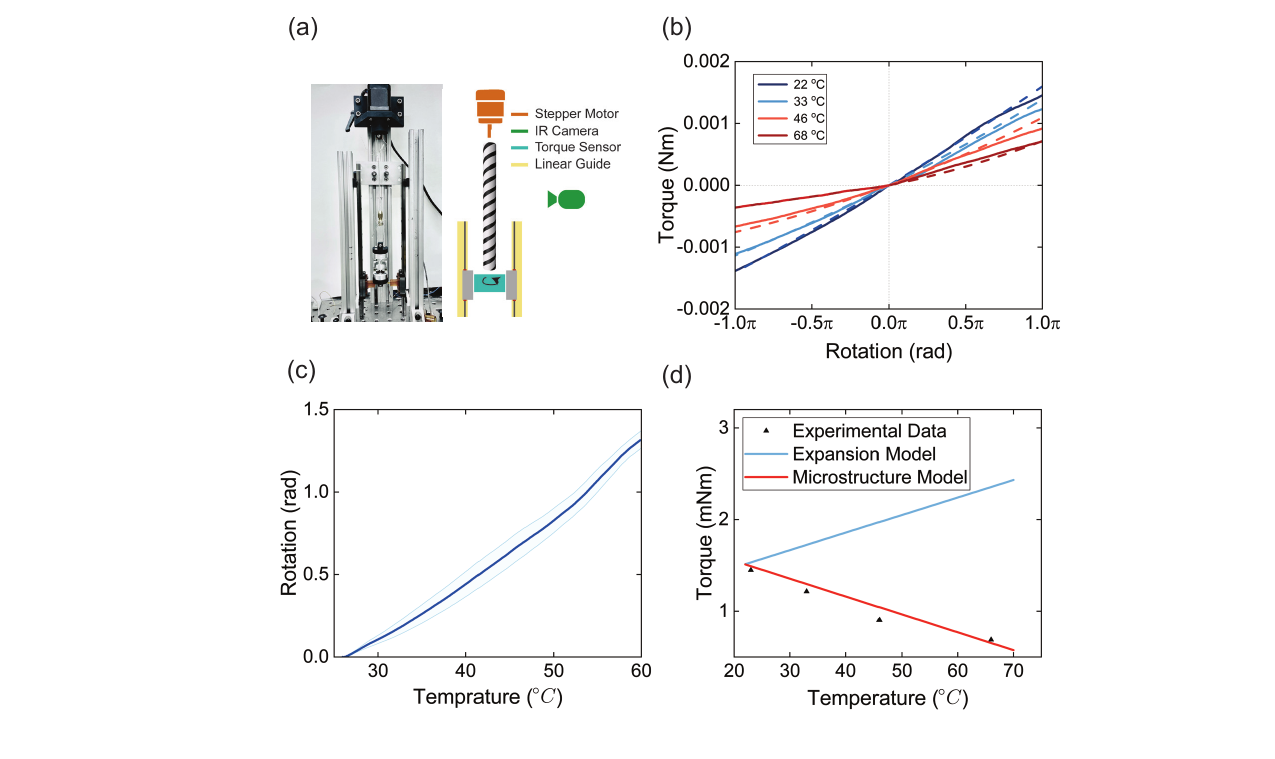}

        \caption{ Twisted and annealed fiber Torsional test:
        (a) Test setup for the thermal-torsional test of the twisted and annealed fiber;
        (b) Plot of the reaction torque generated by the fiber at different temperatures regarding the end rotation. The solid lines stand for the experimental results and the dashed lines stand for the prediction from the microstructure model; The asymmetry with respect to the equilibrium angle is due to the chirality of the twisted and annealed fiber;
        (c) The plot of the end rotation generated by the fiber when no torsional constraint is posed at the bottom. The blue solid line and light blue area - average and standard deviation of experimental results. The red dashed line - microstructure model;  
        d) The plot of temperature versus torque in the case of the expansion model and the microstructure model with experimental data plotted in black markers for reference.}
        \label{fig: TorsionalFiberTest}
\end{figure}

To carry out these investigations, we created a test setup as shown in \cref{fig: TorsionalFiberTest}a. In the fixed-end reaction torque test, the twisted and annealed fiber is attached to the motor that controls the rotational angle and the sensor that measures the reaction torque generated by the fiber upon heating. The axial motion is restricted by the linear guide connected to the torque sensor. In the free-end rotational test, the fiber is heated up with no external torque constraint and the end rotation provided by the fiber is recorded. The temperature in both tests is controlled by the nichrome heating wire wrapped around the fiber and monitored by the infrared (IR) camera. 

The twisted fiber sample is prepared from 1 mm nylon monofilament fiber. The fiber is first twisted and then annealed in oil at 170 $^\circ C$ for 7 hours. During the twisting process, the straight fiber is stretched by a 1 kg constant external load (12.4 MPa) to prevent coiling. The amount of twist inserted into the fiber is 220 turns/m. The sample prepared in this section is almost the same as the sample made in \cref{sec: fabrication} except for the coiled structure. 

 The results of the fixed-end reaction torque test are shown in \cref{fig: TorsionalFiberTest}b. The reaction torque generated by the fiber at different temperatures is plotted for different angular displacements. Four temperatures are selected ranging from room temperature to the typical actuation temperature of TCPAs. As the fiber is heated, the torsional rigidity decreases significantly. In the temperature range where the test is conducted, up to 67 \% change in rigidity is observed as shown in \cref{fig: TorsionalFiberTest} (from around 0.014 Nm to 6e-4 Nm). Moreover, the torsional rigidity of the fiber toward the direction along which the fiber was twisted during manufacturing is slightly larger than that of the opposite direction, which is a sign of chirality. In other words, the test shows that it is easier to 'untwist' the sample than to 'twist' the sample even more.

\cref{fig: TorsionalFiberTest}c illustrates the free-end rotation of the sample when heated without external constraints. As the temperature increases, the fiber provides end rotation toward the untwist direction. This indicates that the sample tends to 'untwist' with regard to its original twist direction when heated up. This procedure is reversible and repeatable under the test conditions. This unique phenomenon is critical to the mechanism of TCPAs. We shall discuss this in more detail in the next section.
        
\subsection{Similarity to Two-way Shape Memory Effect}
\label{subsec: Two-way Shape}

The reversible behavior observed in the \cref{subsec: Thermo-Mech Twisted} is similar to the two-way shape memory effect. In this section, we will discuss the similarity of the TCPA and the two-way shape memory materials from both the manufacturing aspect and actuation mechanism. 

The manufacturing procedure of the TCPA is almost analogous to that of the two-way shape memory polymers (SMP) as shown in \cref{fig: ProgrammingSMPTCPA}a. The stress-temperature-strain curve of programming and actuating a two-way shape memory polymer is shown in \cref{fig: ProgrammingSMPTCPA}b-c. In the two-way SMP, there are generally two structures (Structure A and B) with two distinct thermal transition temperatures ($T_B>T_A$) \cite{Lendlein2019ReprogrammablePolymers,Yarali2020APolymers:}. The polymer is usually stretched at the thermal transition temperature of structure B ($\approx T_{reset}$), leading to a strained and highly oriented polymer chain network. 
Then, the polymer is cooled below the thermal transition temperature with the external force holding the shape as shown in \cref{fig: ProgrammingSMPTCPA}b. If the sample is then cycled between room temperature and $T_g$ (or temperature below $T_m$), only the amorphous region will undergo a significant softening/stiffening effect and a reversible conformational change \cite{Chung2008a,Zare2019Thermally-inducedApplications}. Repeatable actuation is then realized at that specific temperature range just like what we see in two-way shape memory polymer as shown in \cref{fig: ProgrammingSMPTCPA}c which has been attributed to crystalinity-induced elongation during cooling. 

The mechanism of two-way SMP, as briefly mentioned above, is essentially a direct result of stored strain \cite{Baghani2012APolymers,Li2017ARate}, where competition between two components with distinct thermal properties leads to reversible actuation. It can be achieved through different methods, such as liquid crystalline elastomers, copolymers, or interpenetrating networks \cite{Basak2022Two-WayTrends}. Indeed, this concept of changing material properties through mechanical processes is not new to the polymer industries. For almost all polymers, the chemical structure is not the only factor that determines the thermomechanical properties; the morphology and orientation of the polymer chains are also vital. Taking the nylon monofilament fibers used in this work as an example, the cold-drawn process will transform the $\gamma$-crystalline phase to the $\alpha$-crystalline phase, leading to much stiffer fiber ($E_\gamma \approx135 GPa, E_\alpha \approx 295 GPa$ \cite{Li2002}). 

Now we shall take a look at programming and actuation of a nylon TCPA. The precursor fiber is a monofilament nylon commercial fishing line (from \textit{Shaddock Fishing}, 1 mm clear monofilament fishing line, 126.7 lb test strength). We use differential scanning calorimetry (DSC) to characterize the crystallinity and thermal properties with details described in the Appendix B. The material is nylon 6-6,6 copolymer with glass transition temperature $T_g$ at around 45 $^\circ C$ and peak melting temperature $T_m$ at 191.9 $^\circ C$ with initial crystallinity around 23.92 \%. Within the precursor fiber, the ordered crystalline lamellae are surrounded by a disordered amorphous region (\cref{fig: Microstructure}f). As illustrated in \cref{fig: ProgrammingSMPTCPA}a-b, the precursor fiber is first twisted and coiled at room temperature, with both portions highly strained and forced to align toward the twisting direction. At this stage, the fiber has high-strain energy stored within it. Then the polymer is heated to $170^\circ C$, which is at the end of the rubbery plateau and the start of the melting range. The amorphous region becomes rubbery and significant chain mobility is triggered. Their conformation will change to accommodate the applied external strain. After this annealing process, the shape of the fiber is fixed with a reconfiguration of the amorphous region to balance the strain in the crystalline and interphase regions. The crystallinity of the fiber after annealing is around 31.88 \% which is larger than the precursor fiber. If the sample is then cycled between room temperature and $T_g$ (or temperature below $T_m$), only the amorphous region will undergo a significant softening/stiffening effect and conformational change that is reversible. Repeatable actuation is then realized at that specific temperature range just like what we see in two-way shape memory polymer as shown in \cref{fig: ProgrammingSMPTCPA}c. Although there are similarities in the actuation of the two-way shape memory polymer and TCPA, it's important to note that they are not identical. The actuation of the two-way shape memory polymer is usually linked with a specific actuation around the melting point of the semi-crystalline polymers. Conversely, the actuation range for TCPA is very broad as it begins beneath the glass transition temperature and continues seamlessly until around 20 °C prior to the onset of the melting peak.

As we have seen, the shapes of both the SMP and nylon TCPA are fixed by heating the sample to the thermal transition temperature. In two-way SMP, the samples are highly deformed, usually axially after heated to programming temperature. In the programming processes of nylon TCPA, the samples are highly strained torsionally at room temperature creating normal strains of up to 60\% and then heated to the programming temperature. The only major difference is the sequence of loading and heating as indicated in \cref{fig: ProgrammingSMPTCPA}a. 

\section{Complete Mathematical Model for TCPA}
\label{sec: Complete Math Model}

From \cref{sec: Love Model}, we have obtain the model for the twisted helical geometry of the TCPA. However, this model does not include the effect of twisting and annealing of the Nylon fiber. In this section, we will expand this model to the case of TCPAs by introducing the coupling between the bending mode and twisting mode and the micro-structure model. We will first introduce the coupling term which accounts for the non-isotropic characteristic of the twisted coils. Then based on the experimental results obtained in \cref{sec: Thermo-Mech Test}, we construct the mathematical microstructure model to account for the effect of annealing during manufacturing. By combing the coupling term and the microstructure model into the rod theory, we derive the complete mathematical expression of the potential energy of the system, which can then be used to find the equilibrium configuration.

\subsection{Modified Kirchhoff Rod with Coupling Between Bend and Twist} 
\label{subsec: Couple}

\begin{figure}[ht]
\includegraphics[width=1\textwidth]{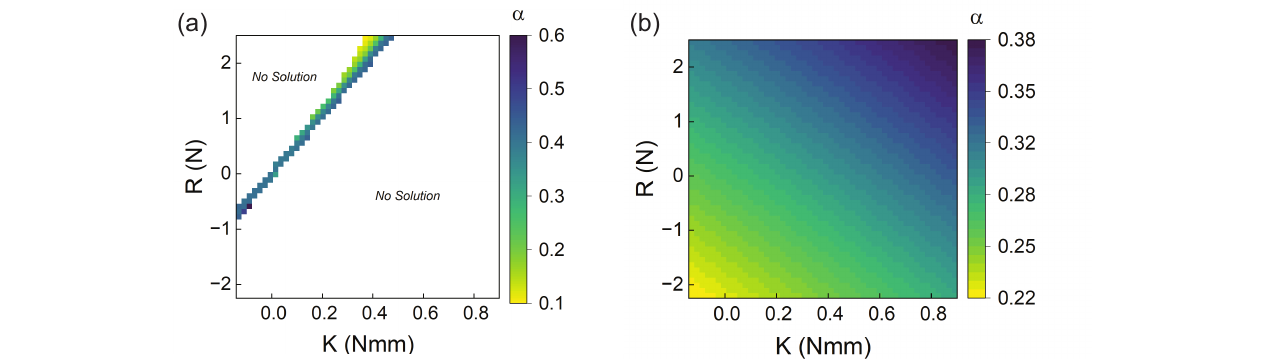}

        \caption{  Prediction and validation of the full model.
(a) Contour plot of the equilibrium solutions to the full system at room temperature without coupling, the white area indicates that there is no helical solution for this combination of external force and torque; (b) Contour plot of the equilibrium solutions to the full system at room temperature with coupling which shows that solution exists for arbitrary combinations of external forces and torques.
}
        \label{fig: CouplingTerm}
\end{figure}
In this subsection, we will modify the strain energy from classic Kirchhoff Rod by adding a coupling term between bend and twist required to model TCPAs. In the formulation of elastic energy proposed in \cref{subsec: Strain Energy}, one can only find two steady helical solutions, one right-handed and one left-handed. This indicates that for a certain helix angle $\alpha $, a certain combination of the external force and torque needed to be satisfied to obtain equilibrium. Clearly, this is not the case for the annealed nylon TCPA, which exhibits equilibrium at any combination of force and torque. This implies that the isotropic assumption we have made by considering only the bending and torsional rigidity causes limitations in the utility of the model. Here, we draw inspiration from the mathematical modeling of DNA. Both DNAs and TCPAs can be conceptualized as 'twisted helices', albeit on different scales and with distinct materials. DNA and TCPAs share two important features in common. The first is that there is a significant contrast between the mechanical properties along the axial direction and the radial direction, leading to anisotropic materials. The second is that the fibers are torsionally constrained, making them twist-storing polymers \cite{Moroz1998EntropicPolymers}. Marko and the following researchers \cite{Marko1994,Nomidis2017,Nomidis2019} have shown that the bending and torsional mode of the double-stranded B-DNA double helix is not independent of each other. They proposed that due to the geometrical chirality induced by the double helix structure, one shall see that twist-bend coupling contributed to the free energy of the system. 

 TCPA samples are made from highly aligned monofilament fibers. They have a considerably high anisotropy ratio of radial to axial direction. Thus we assume that the coupling between bend and twist is non-zero, leading to $C_{13}=C_{23}\neq 0$. The fibers are still radially isotropic, giving $B_1=B_2=B$. Given that the geometric torsion $ \tau_s$ typically is one order of magnitude smaller than the internal twist $\tau_i$ within the context of TCPA, it is reasonable to neglect the former component. Consequently, the primary interaction is anticipated between the internal twist $\tau_i$ and the bending (curvature $\kappa$). Following the above assumption, we can reformulate the strain energy in \cref{eqn: ElasEnergyLong} related to the coupling between bend and twist as
\begin{align}
    & U= U_B+U_T+U_C,\quad    U_C= S B_C (\kappa - \kappa_0) (\tau_i- \tau_{i0}) \label{eqn: coupleEnergy},
\end{align}
where $B_C$ is the measure of the intensity of the coupling. To illustrate the effect of the coupling term, we present two figures: \cref{fig: CouplingTerm}a depicts the solution of the system without considering the coupling effect, while \cref{fig: CouplingTerm}d demonstrates the solution with the incorporation of the coupling term. Evidently, the introduction of the coupling term enables solutions across a wide spectrum of loading conditions, while the absence of the coupling poses a strict constraint on the permissible external loading.

\subsection{Mathematical Microstructure Model}
\label{subsec: Math Micro Model}

\begin{figure}[ht]
\includegraphics[width=1\textwidth]{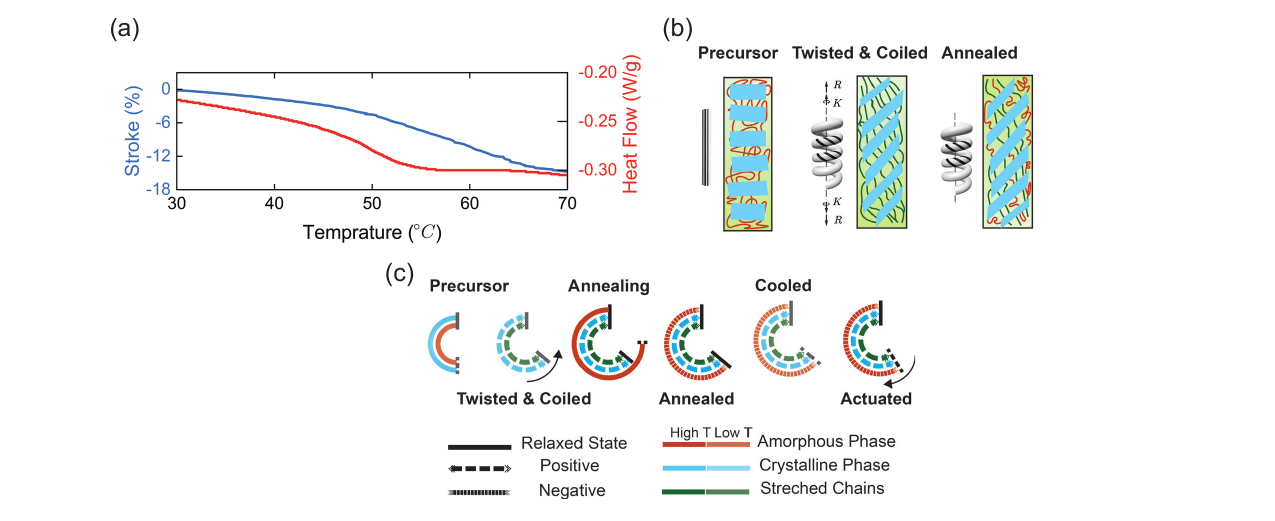}

        \caption{Mathematical microstructure model:
        (a) Displacement curve of the TCPA under free heating (blue) and the DSC curve of the sample (red);
(b) Schematic of internal structure for precursor fiber, twisted and coiled fiber, and annealed fiber, which is three representative stages during the manufacturing of the Nylon TCPA. The blue rectangles indicate the crystalline phase, the brown solid lines indicate the amorphous phase and the green solid lines are stretched chains;
(c) Detailed schematic of the microstructure model at different stages of manufacturing and actuation with different line styles referring to different states. The direction of deformation is represented by the line style. Different phases of the material are represented by different colors and the temperature is represented by the shade of the color.
      }
        \label{fig: Microstructure}
\end{figure}

In this subsection, a microstructure model is proposed based on our analysis in \cref{subsec: Two-way Shape}. We take inspiration from the classical Takayanagi model \cite{Takayanagi2007} as well as the models for SMP \cite{Chen2008ADeformations,Liu2006ThermomechanicsModeling, Li2017ARate,Nguyen2008ARelaxation, Bhattacharyya2000AnalysisModel} and modify it to accommodate the case discussed here. Takayanagi proposed a composite model composed of AC parallel and AC series connections where A stands for amorphous and C stands for crystalline. This model and its alternative forms have been widely used to simulate the thermal transport of semicrystalline materials \cite{Lu2018b}. On the other hand, researchers have proposed a two-phase model (e.g. frozen phase and active phase) to explain the mechanism of the shape memory polymers \cite{Chen2008ADeformations}. The phases are commonly modeled as springs in series or in parallel which store strain and release them upon activation \cite{Nguyen2008ARelaxation,Li2017ARate}. Here, we adopt the parallel configuration since it is the primary factor that enables the actuation as we will see later. In addition, we change the linear spring in the Takayanagi model to the torsional spring since the primary mode in TCPAs is the torsional mode. As indicated in \cref{fig: Microstructure}c, the fiber is simplified into two connected parallel torsional springs. 

The red spring represents the amorphous phase and the blue spring represents the crystalline phase. After twisting and coiling, the chains in both phases are highly stretched toward the positive direction. We denoted the stretched amorphous phase as stretched chains. During annealing, part of the stretched amorphous phase changes its conformation and releases the strain energy. Upon cooling, the released chains and stretched chains form a new equilibrium where no external force is needed to hold the structure. In the model, these released chains are represented by the amorphous phase spring that is compressed toward the negative direction. If the structure is heated, the modulus of the amorphous phase will decrease much faster than the modulus of the crystalline phase. The stretched amorphous phase and the crystalline phase spring will then pull equilibrium counter-clockwise, leading to a torsional actuation. 

While we have been using the linear spring to describe the mechanism of twisted fiber samples, one factor also needs to be included in the model. As shown in \cref{fig: Microstructure}b, the reaction torque generated by the fiber is different along the positive and negative directions. The fiber is stiffer towards the positive (twist) direction and softer towards the negative (untwist) direction. This effect can be reflected by including a second-order stain term. For a straight fiber, only the torsional mode is considered and no external force is needed to hold it. Thus one can formulate the torsional energy $\overline{U}_T$ as
\begin{align}
    \overline{U}_{T}=(\frac{1}{2}(C(\tau-\tau_p)^2+C_{am}(\tau-\tau_0)^2)+\frac{1}{3}C_3 (\tau-\tau_{00})^3)S, \label{eqn: FiberTorEnergy}
\end{align}
where $\tau_p=\sin{\alpha_0}/l_0$ is the initial geometry torsion,  $\tau_0$ stands for the extent of relaxation for the amorphous phase, and $\tau_{00}$ is the zero shift of the higher-order term. These constants' shifts are incorporated to accommodate plastic deformation and annealing processes during manufacturing which changes the material structure permanently. Here, we treat these effects as a unified entity, evaluating them by measuring the static geometry of the actuator after all manufacturing procedures, without distinguishing between the two. $C$ is the torsional rigidity of the stretched crystalline and amorphous region, $C_{am}$ is the torsional rigidity of the relaxed amorphous region and $C_3$ is the higher-order elastic constant. With the condition $\kappa=0,\tau=\tau_{i0}$, one shall derive the potential energy of a straight and annealed fiber as
\begin{align}
    \overline{\Pi}=\overline{U}_T+W_K=(\frac{1}{2}(C(\tau-\tau_p)^2+C_{am}(\tau-\tau_0)^2)+\frac{1}{3}C_3 (\tau-\tau_{00})^3-K(\tau_i-\tau_{i0}))S.
\end{align}

Here the only free parameter that defines the structure of the fiber is the internal twist $\tau_i$. The equilibrium equation is given by taking the derivative of \cref{eqn: FiberTorEnergy} with respect to $\tau_i$:
\begin{align}
     \frac{\partial\overline{\Pi}}{\partial \tau_i}=0   \to K=C(\tau_i-\tau_p)+C_{am}(\tau_i-\tau_0)+C_3 (\tau_i-\tau_{00})^2. \label{eqn: TwistedFiber}
\end{align}

\cref{eqn: TwistedFiber} gives the relation between internal twist and external torque and can be fitted into the thermomechanical test in \cref{fig: TorsionalFiberTest}b-d. It is assumed that the linear torsional rigidity $C_{am}$ changes linearly with temperature, while $C$ remains constant. The rigidities are given by the total rigidity and the crystallinity $C^0=C^0_{tot} \Lambda$ where the crystallinity $\Lambda\approx0.3$ is determined through the DSC test. The internal twist $\tau_i$ can be written as a function of the end rotation $\Omega$, fiber length $L$ and the initial internal twisted $\tau_{00}$: $\tau_i=\frac{\Omega}{L}+\tau_{00}$. Then \cref{eqn: TwistedFiber} can be reorganized as
\begin{align}
     &K=C (\frac{\Omega}{L}+\tau_{00})+C_{am}(\frac{\Omega}{L}-\tau_0+\tau_{00})+C_3 (\frac{\Omega}{L})^2, \\
     C_{am}=C_{am}^0 & + \zeta_C T  ,C=C^0, C_{tot}^0=C^0+C^0_{am},C_3=C_3^0+\zeta_{C3} T. \label{eqn: defC}
\end{align}
     
The reaction torque and free-end rotation provided by the twisted fiber can then be predicted as the dashed lines shown in \cref{fig: TorsionalFiberTest}b-c. The parameters used in the fitting are listed in \cref{table: Torsional Test}. The results show good agreement with the experimental data and we shall use this set of data in the following section for the torsional rigidity of the material. In \cref{fig: Microstructure}d, the comparison is made between the model proposed and the expansion model based on radial swelling of the fiber. In the expansion model, only the radius of the fiber is considered to be changing upon heating. By approximating the torsional rigidity by $C=GJ$, the torque generated by the fiber will increase with temperature, since the polar momentum of the rod increases. This is inverse to what we observed in the actual experiments and indicates that the proposed model aligns better with the behavior of the twist fiber sample.

\subsection{Complete Model of TCPA}
\label{subsec: Complete model}
Now that the microstructure model shows its ability to predict the performance of a straight twisted and annealed fiber, and the bend-twist coupling term expanded the equilibrium possibilities of the twisted helix, we are ready to present the full TCPA model. In \cref{subsec: Math Micro Model} and \cref{subsec: Couple}, we have discussed how the classic Kirchhoff rod theory can be modified to accommodate the unique characteristics of the TCPAs. By combining the microstructure model with the macroscopic model, the complete system can be constructed to describe a TCPA during actuation. The total potential energy is updated to 
\begin{align}
\Pi & = U+W=(U_B+\overline{U}_T+\overline{U}_C )+(W_K+W_R), \label{eqn: CompletePotential}
\end{align}
where the work potential $W$ is given by \cref{eqn: WorkPotential}, and the elastic energy is given by \cref{eqn: FiberTorEnergy} and \cref{eqn: coupleEnergy}. The equilibrium solution of the TCPA at a given environmental and loading condition can then be solved according to \cref{eqn: PartialCond}. In order to validate the assumption we have posed about the torsional energy $\overline{U}_T$ and coupling energy $\overline{U}_C$, several different variations are adopted to showcase how different formulations can influence the results. 

To study how the expression of the torsional energy influences the results, we investigate three different variations of the $\overline{U}_T$. The first case (case A) is the single-phase microstructure model where we assume that there is only one phase in the material. The second case (case B) is the two-phase microstructure model where there are two different phases in the material and only the quadratic terms are taken into consideration. The final case (case C) is the high-order two-phase microstructure model which has been proposed in \cref{eqn: FiberTorEnergy}:
    \begin{gather}
        \textbf{Case A} - \overline{U}_T= \frac{1}{2} S C(\tau-\tau_0)^2 \label{eqn: TorsionEnergyA},\\
        \textbf{Case B} - \overline{U}_T= \frac{1}{2} S (C(\tau-\tau_p)^2+C_{am} (\tau-\tau_0)^2),\\
        \textbf{Case C} - \overline{U}_T= \frac{1}{2} S (C(\tau-\tau_p)^2+C_{am} (\tau-\tau_0)^2)+\frac{1}{3} S C_3 (\tau-\tau_{00})^3. \label{eqn: TosionEnergyC}
    \end{gather}        

Similarly, we study how the coupling influences the results by investigating the following two variations of the coupling energy term proposed in \cref{eqn: coupleEnergy}. The first case (case 1) is the coupling term proposed in \cref{eqn: coupleEnergy}. The second formulation (case 2) considers the case where there exists a higher-order term in coupling:
\begin{gather}
         \textbf{Case 1} - \overline{U}_C=B_{C1} S (\tau_i-\tau_{i1})(\kappa-\kappa_p),\\
       \textbf{Case 2} - \overline{U}_C=B_{C1} S (\tau_i-\tau_{i1})(\kappa-\kappa_p)+B_{C2} S (\tau_i-\tau_{i2})(\kappa-\kappa_p)^2. \label{eqn: CoupleEnergy2}
\end{gather}

\section{Model Validation}
\label{sec: Model validation}

In this section, the proposed model is verified through two tests. The DMA test represents conditions of slow heating while the water isobaric test represents the case of fast heating. In real-world applications, TCPAs are typically subjected to heating rates that fall between our two validation scenarios. Due to viscoelastic relaxation, the heating rate significantly impacts TCPA performance. While the proposed model does not consider the viscoelasticity, it captures the behavior both at a low heating rate and a high heating rate based on the corresponding softening parameters characterized experimentally. The prediction of the model shows good agreement with the experimental results and proved its ability to predict TCPA performance under various conditions.

\subsection{Validation of TCPA Actuation Measured in the DMA}
\label{subsec: DMA validation}

In this section, we will use the DMA test results of the TCPA to validate the model. The procedure of the DMA test is described in \cref{fig: DMATest}b with the free heat-up process (step b1). The stress-strain response of the TCPA sample at room temperature and actuated temperature is plotted separately in \cref{fig: Prediction}a-b by a solid black line. Consider all the cases described in \cref{subsec: Complete model}, the potential energy can be constructed with six different combinations. All the models and material parameters involved in the simulation are listed in \cref{table: DMA}. During the annealing, the stress induced by the internal twist can be released, and it is not possible to determine its value directly from the experimental measurement. Thus, the initial internal twist $\tau_{00}$ remains a fitting parameter. The initial values of the twist and curvature are given by the geometry $\kappa=\frac{cos \alpha_0}{l_0}, \tau=\tau_{00}+\frac{sin \alpha_0}{l_0}$. Parameters $\{\tau_0,\tau_p,\tau_{i1},\kappa_p\}$ can be directly derived under the condition that the sample is initially at rest and that no external load is needed to hold its structure. In terms of material properties, the constants of the coupling terms ($ B_{C1}, B_{C2},\tau_{i2}$) are optimized with Optuna \cite{Akiba2019Optuna:Framework} for each formulation. The predictions of both non-actuated and actuated states resulting from these six cases are presented in \cref{fig: Prediction}a-b. The error of each prediction is depicted in \cref{fig: Prediction}c by taking the summation of the mean squared error of the predicted strain value compared to the experimental strain for both non-actuated and actuated states.  

The results confirm that the model proposed in this work, case \textbf{B2}, provides the most accurate predictions for the performance of the TCPA. This underscores the importance of both the dual-phase microstructure model and the coupling to model the behavior of TCPA. Among these factors, the dual-phase micro-structural model is the most crucial component, since only cases \textbf{B} and \textbf{C} exhibit active stroke. Without this assumption, the TCPA would lose contrast between the two phases and elongate upon heating. The higher-order torsional term has minimal influence on the system, as there is little difference observed between cases \textbf{B} and \textbf{C}. However, the inclusion of a higher-order term in the coupling, as seen in case \textbf{2}, improves the prediction. This prediction's precision is on par with other suggested models, whose accuracy lies approximately within the 5\% to 15\% range \cite{Yang2016b,Hunt2021Thermo-mechanicalFEA,Wu2020} but with the added advantage of capturing the annealed shape via an analytical framework.

\subsection{Validation of the TCPA Actuation from a Fast Isobaric Test}
\label{subsec: Isobaric Validation}
In this section, we will validate the proposed model from a fast-heating untethered water isobaric test. It is commonly observed that the performance of the TCPA is greatly influenced by the speed of heating. Nylon, as a semi-crystalline polymer, exhibits different changes under varying heating rates. When subjected to higher heating rates, the polymer chains have less time to rearrange, resulting in a delayed softening effect of the fiber. This phenomenon is particularly pronounced when the TCPA is actuated by immersion in water, where the heating rate is approximately 41 $^\circ C/s$. This rate is almost 270 times higher than that observed in the DMA test (0.15 $^\circ C/s$) discussed in the previous section. Consequently, the reduced softening effect leads to a significant increase in the active actuation provided by the actuator. Also, while nylon is known to swell with water, a very fast immersion test (few seconds) minimizes this effect, and it can be neglected. 

To investigate the thermomechanical properties of nylon fiber under fast heating, a specialized setup was designed, as depicted in \cref{fig: WaterTestSetup}. The test relies on fast heating by immersion in hot water. The tested sample was prepared by twisting and annealing the straight nylon fiber, following the procedure outlined in \cref{sec: Role of Micro}. The experimental procedure of the torsional test is outlined in the Appendix. The test results indicate that the reaction torque of the fiber decreases upon immersion in water as shown in \cref{fig: WaterTestSetup}. This is consistent with the slow heating test. Through the new tests, we acquired a new set of thermomechanical parameters of the samples, as detailed in \cref{table: water}. While the model fits well the results of the water test, it uses different numerical values for the parameters as can be observed by comparing the parameters in \cref{table: DMA} and \cref{table: water}, which is due to the different heating rate.

\begin{figure}[t]
    \centering
    \includegraphics[width=1\textwidth]{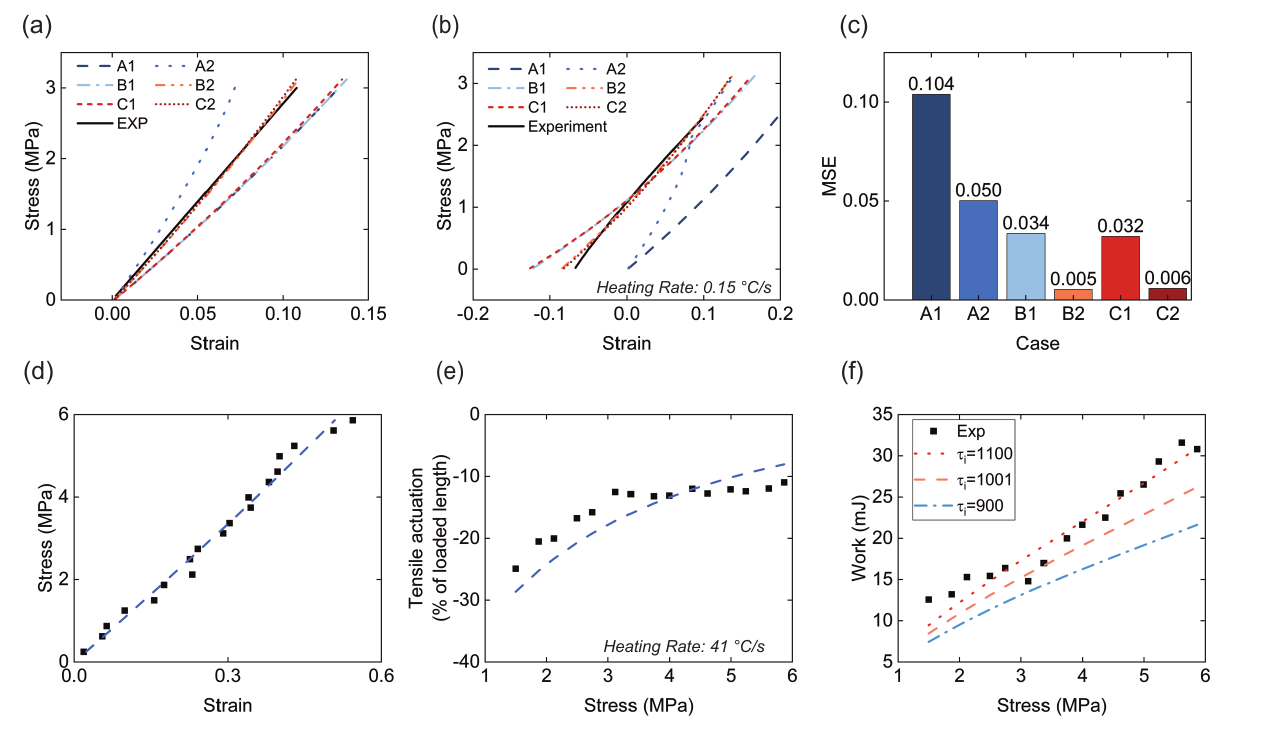}
    
    \caption{Experimental validation of the proposed model: (a-b) The plot of the experimentally measured and mathematically modeled tensile response at (a) non-actuated state (room temperature, 25 °C) and (b) actuated state (actuation temperature, $\approx$ 65 °C). The black line stands for the experimental results from the DMA test (where the heating is slow) while the colored lines stand for model predictions from different formulations of the potential energy. The stress is normalized by the cross-sectional area of the fiber, and the strain is normalized by the original length of the muscle before testing;
    (c) The plot of the sum of the mean squared error of the predicted performance for both actuated state and tensile response;
(d-e) The prediction of the tensile strain response and actuation performance from the model in dashed line and the experimental results (where the heating is fast) in black markers from the water isobaric test;
    (f) Work provided by the actuation for varied internal twists from model prediction compared to the experimental result of the water isobaric test.}
    \label{fig: Prediction}
\end{figure}

Incorporating the parameters in \cref{table: water} into the system, we project the anticipated performance of the muscle under the isobaric testing condition, as depicted in \cref{fig: Prediction}d-e. The prediction aligns closely with experimental results. It is worth noting that the model falls short of precisely representing the nonlinearity of the actuation performance. The work carried out by the actuator during actuation is plotted in \cref{fig: Prediction}f. Both experimental data and the model prediction with varied internal twists are shown in the plot. As the initial degree of internal twist increases, the model suggests that the actuator performs a greater amount of work, primarily attributed to the elevated level of stored energy which will be discussed in the next section.

\section{Discussion}
\label{sec: Discussion}

From \cref{sec: Model validation}, we have validated the proposed model through the experimental results. In this section, we proceed to analyze some important behavior of this system and derive some important insights. We will start with the advantages of this model which is the continuous prediction for both passive and active states. Then we will present the prediction provided by the model of how the geometry and material selection can potentially enhance the performance of the TCPAs. Finally, we will discuss the limitation of this model and how it can be address in the future.

\subsection{Modeling the Elasticity of the Passive and Active States}
This study emphasizes the importance of studying the elasticity of the actuator in the active state, whereby softening is observed to occur and will affect the performance. This softening of the actuated state cannot be observed from the standard isobaric test. Previous models have focused on the tensile actuation stroke of actuators at a given constant load \cite{Sharafi2015AMuscles, Yang2016b}. Unfortunately, the force-displacement relationships of the active and passive states are often ignored in the model validation. Yet, the stiffness of an elastic actuator is as important as other metrics, such as stroke, load, and work capacities, for performance prediction when integrated into a mechanical system. In real-world applications, actuators interact with external loads or are embedded within elastic structures. During their operation, the elasticity of actuators, i.e. their force-displacement relationship, changes between the active and passive states, leading to different behaviors under different loading conditions. In this work, we include not only the actuation performance as shown in \cref{fig: Prediction}b,e, but also the tensile test response as shown in \cref{fig: Prediction}a,d. The prediction of the tensile response is valuable for real-world applications, such as recent jump robots powered by TCPAs \cite{Wang2023Insect-scaleCascade}. In that case, the TCPA pulls a curved beam to buckle it and the loading curve is highly non-linear. The trade-off between softening and contraction influences the stroke provided by the actuator. At higher loads, the contraction decreases since the external force stretches the muscle more as it becomes softer. The stroke as a function of stress, called isobaric performance, is insufficient to predict the actuator performance. The passive and active response predictions from the proposed model shown in \cref{fig: Prediction} are equally needed for a complete prediction of the actuator performance under arbitrary loading conditions.

\begin{figure}[ht]
\includegraphics[width=1\textwidth]{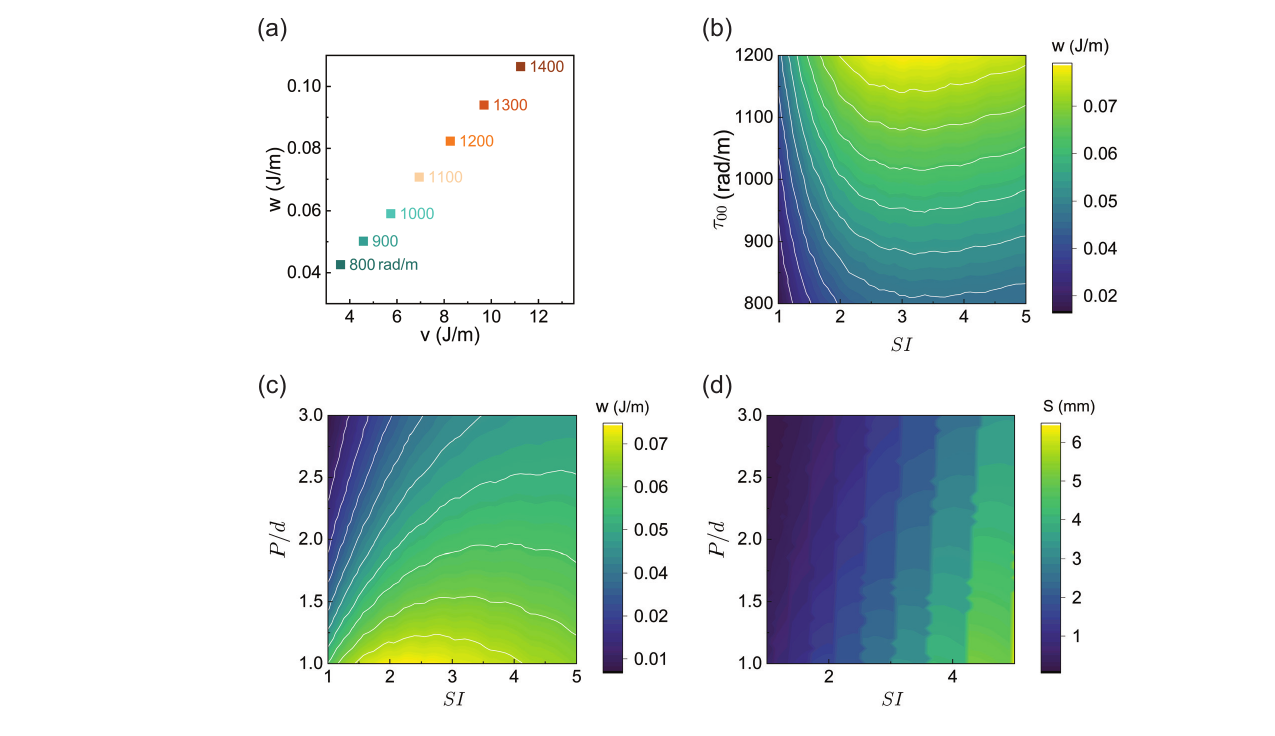}
        \caption{ (a) Actuation work per fiber length \textit{w} versus the stored elastic energy per fiber length \textit{v} where a positive correlation between the two is observed. The marker inside the figure indicates the initial internal twist of the rod;
(b) Contour plot of the maximum actuation work per fiber length \textit{w} with spring index \textit{SI}, initial internal twist $\tau_{00}$. An optimized value of spring index is observed for a constant internal twist value;
(c)Contour plot of the maximum actuation work per fiber length \textit{w} with normalized pitch $P/d$ and spring index \textit{SI} where larger pitch leads to lower actuation work;
(d) Contour plot of the stroke corresponding to maximum work for different values of $P/d$ and \textit{SI} which show that the stroke at maximum work is primarily dependent on the spring index and less dependent on the normalized pitch. 
}
        \label{fig: Discussion}
\end{figure}

\subsection{Performance of TCPA as a Function of Internal Twist and Geometry}
\label{subsec: Improve}
As explained by the proposed theory, TCPAs actuate by releasing the stored elastic energy of the highly strained structure. Thus, the actuation performance of the TCPA depends on the amount of stored elastic energy and the capability of the structure to release that energy and convert it into useful mechanical work. The level of the stored energy is proportional to the amount of the initial twist inserted into the fiber since the fiber is mainly twisted during the manufacturing process. \cref{fig: Discussion}a shows the relationship between the actuation work per fiber length and the elastic energy per fiber length of the actuator with the labels showing the corresponding initial twist $\tau_{00}$. With a larger initial twist, the stored elastic energy per fiber length increases. In that case, more elastic energy is available for use in the actuation process, which leads to a larger work per fiber length. This leads to the primary approach of improving the performance, which is to store more energy in the material during the manufacturing process. However, in practice, the increase of the initial twist is limited by the toughness of the material. Straining the fiber beyond its ultimate strength can cause fracture during the twisting or annealing process. This explains why fishing lines, which are strong and tough, are suitable for coiled artificial muscles.

The second approach to increase the work is to improve the capability of the structure to release energy. Insights into the capability of the muscle to do work by releasing the stored elastic energy is gained by examining the relationship between the x and y values of \cref{fig: Discussion}a. This capability depends on the parameters of the system such as the geometry. The analysis is performed on the system described in \cref{subsec: DMA validation} with a fixed fiber length. There are different ways to measure the actuation performance such as maximum stroke, maximum force, and maximum work. Among them, the maximum work of the actuator is a more complete measure of performance since it contains both information about stroke and stress. So, for each muscle geometry, we scan a range of applied loads and their corresponding strokes, and select the combination that maximizes the work and plot that value. \cref{fig: Discussion}b shows the influence of the initial twist $\tau_{00}$ and the spring index \textit{SI} (helix diameter over fiber diameter) on the maximum work of the system. As the initial internal twist increases, the maximum actuation work per fiber length increases, confirming the results obtained in \cref{fig: Discussion}a. For each $\tau_{00}$, an optimum actuation work is observed at an intermediate \textit{SI}. \cref{fig: Discussion}c shows how the pitch \textendash normalized by the fiber diameter $P/d$ \textendash and \textit{SI} change the behavior of the system. Again, at each value of $P/d$, a maximum actuation work per fiber length is achieved at an intermediate value of \textit{SI}. Also, a higher value of maximum work is obtained for smaller pitch muscles. This analysis indicates that the performance of the TCPA can be improved by finding the minimal pitch and optimal \textit{SI} of the helical structure. \cref{fig: Discussion}d shows the stroke corresponding to the maximum work in \cref{fig: Discussion}c. The normalized pitch $P/d$ does not influence the stroke provided by the actuator significantly. The stroke is predominantly dependent on \textit{SI}, whereby a small value of \textit{SI} corresponds to a small stroke. However, as \textit{SI} further increases, the actuator becomes softer, which leads to lower operational stress. Thus the work starts to decrease when this softening effect becomes dominant over the increase in stroke.

In conclusion, the performance of TCPA can be improved by both the internal twist and the geometry. Compared to increasing the internal twist, changing the geometry such as decreasing pitch and finding the optimum spring index value is overall easier since it does not require a change in material properties. It requires only a change in manufacturing procedures, which facilitates tunable actuation of actuators using specified materials.

\subsection{New Materials for High Work Capacity Actuators}

The model proposed in this work opens opportunities to use new materials for better soft actuators. While current studies on TCPAs focus on materials with anisotropic swelling behavior such as CNT yarns and nylon fibers, the model proposed in this work proposes exploring energy storage material for effective TCPA actuation. These materials feature components with distinct thermal and mechanical properties, along with high ultimate strain, akin to the prerequisites for shape memory polymers. Three criteria need to be satisfied for these new polymers to make them suitable for manufacture of TCPAs. First, at least two networks need to exist within the material, which has been the case for two-way SMP. Secondly, the material must be tough and ductile to withstand the large strain applied during twisting. Thirdly, the material must be processable into a slender rod for twist insertion. Some early attempts have been published for twisting and coiling SMA, SMP and LCE, such as in \cite{Cui2021,Jiang2021,Hu2023BioinspiredActuators,Lugger2023Melt-ExtrudedActuators,zhang_compound_2024}, showing the potential of this structure. However, there are still challenges in fabricating high-strength fibers using these materials \cite{Lugger2023Melt-ExtrudedActuators}, which can be addressed in future work. 

\subsection{Limitations of the Mathematical Model}

 The model proposed is kept as simple as possible to minimize computation costs and gain physical insight. It has several limitations that could be addressed in further work. The first limitation is that the model assumes that the TCPA is a perfect helix, for which we calculate the equilibrium based on a constant helix angle throughout the center line. This gives the classic helical equilibrium solution of rods under a terminal wrench. However, during the manufacturing process of TCPA, the fiber undergoes twisting until a localized secondary instability kicks in. At this point, one helical turn is formed at a time whereby the fiber snaps up and comes into self-contact within a single coiled turn. As more twists are inserted, more coils are added consecutively. Consequently, the model cannot be used to explain the dynamics of coiling during the manufacturing process. However, it is still valid for the prediction of the post-manufacture actuation performance.

The second limitation is that we did not study the torque required to hold the TCPA during passive or active actuation. The rod needs a certain combination of external force and torque to remain in the helical configuration. Previous work has shown the complications caused by the coupling term with the response of a twisted helix \cite{CrossedDSignurickovic2013TwistFilaments}. It has been shown that for a twisted helix or an engineering spring, the torque needed to hold them during stretching changes non-monotonically with displacement. Since our major focus is on tensile actuation, we will reserve the discussion about this topic for future work. This will require also a new experimental setup to measure this torque during contractile actuation.

The third limitation is that the model does not consider the complex behavior of the material such as viscoelasticity, effect of tie-molecules, and swelling. It is assumed that the actuation of TCPA is a quasistatic process. However, due to viscoelasticity, this assumption is not precise. As one heats up the sample at different rates, different responses are observed for TCPA. This can be an important factor to study especially if the TCPA relaxes or swells such that its viscoelastic material behavior affects its performance. The effect of the tie molecules is also not modeled in the microstructure model. Finally, due to the nature of the Kirchhoff rod where the rod is reduced to the centerline and frames, the swelling-induced actuation can not be directly derived which limits the use of this model.

Finally, the experiments validated the relationship between the actuation performance of the TCPA, its geometric parameters, and its material properties (measured via DMA and torsional tests). We have not addressed the relation between the manufacturing load, twist density, and the resulting coil geometry due to the buckling instability, which is an interesting mechanics problem recently studied for the case of hyperelastic filaments \cite{LiuJMPS}. An extension of this recent study to nylon TCPA should consider the microscopic change during the annealing process.

\section{Conclusion and Outlook}
\label{sec: Conclusion}

In this work, we experimentally measured and mathematically modeled the actuation behavior of nylon TCPA. An analytical model grounded in the classical Kirchhoff rod theory with ad-hoc modifications is used to show how the change in the equilibrium state enables actuation. The model establishes a pragmatic and lightweight theoretical framework that holds the potential to guide the formulation of novel materials and innovative actuator structures. We demonstrate the need for this model by experimentally showing the softening of TCPA in the active state during heating, which was not previously reported.

The model describes the geometry aspect of the TCPA with the Kirchhoff rod theory and the minimum total potential energy principle. We also discussed the significant similarity in the actuation mechanisms between two-way shape memory polymers and TCPAs, which motivates the development of a two-phase microstructure model. The proposed microstructure model is based on the Takayanagi model and inspiration from SMP modeling. A two-phase model is constructed that comprises crystalline and amorphous regions as the two phases that enable elastic energy storage and residual stresses. Combining the microstructure model and the geometric model, we are able to construct a full predictive model for the TCPA. The coupling between bending and twisting is taken into account in the full model. 

A comprehensive series of experiments is conducted to validate the proposed model, namely DMA tests and fast heating and cooling isobaric tests in water medium. It successfully predicts untethered actuation and the softening effect of TCPAs in both tests. While the thermal expansion effect is not captured in this study for simplicity, it can be integrated into the model as an additional term in the strain vectors with which both swelling-based and microstructure-based actuation can be simulated. The model could potentially be extended to the Cosserat rod model and implemented in an FEA framework. Furthermore, the proposed mechanism offers fresh insight into conceiving new material microstructure to produce the next generation of powerful actuators.

\section{Acknowledgment}
We acknowledge funding from the following sources: Defense Advanced Research Projects Agency DARPA SHRIMP HR001119C0042 (Q.W. and S.T.), Toyota Research Institute North America TMNA (Q.W. and S.T.), Office of Naval Research N00014-22-1-2569 (L.C. and S.T.). Grainger College of Engineering Strategic Research Initiative (Q.W. and S.T.).

\section{Declaration of Interests}
The authors declare that they have no known competing financial interests or personal relationships that could have appeared to influence the work reported in this paper.

\bibliographystyle{elsarticle-num} 
\bibliography{TCPATheoryRef.bib}

\appendix
\newpage
\newpage\setcounter{table}{0}
\newpage\setcounter{figure}{0}
\section{Geometry of the Helix}
\label{appen: Geometry}

\begin{figure}[ht]
    \centering
    \includegraphics[width=1\textwidth]{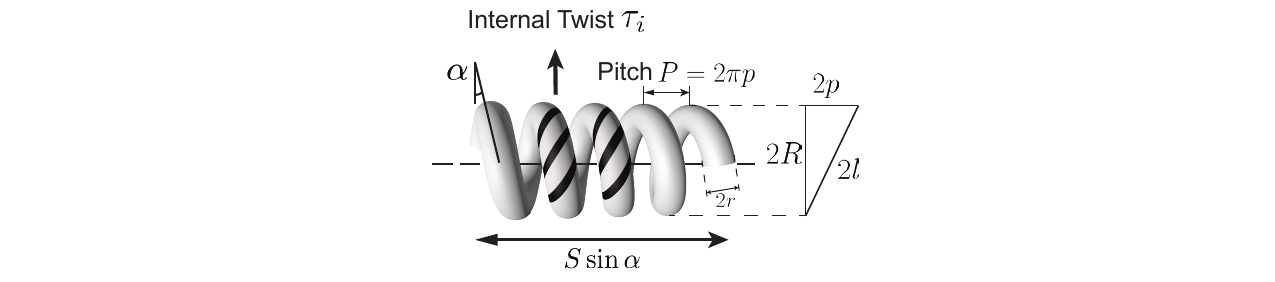}
    
    \caption{Geometry and notation of a twisted helix
     }
    \label{fig: Geometry}
\end{figure}

A twisted helical rod can be described by different geometry parameters. In different theoretical frameworks, we have seen different combinations of these parameters being adopted. Here we will provide a short description of these parameters.

The first set of parameters is commonly used to describe an engineering spring: wire diameter $d$, mean coil diameter \textit{D} and pitch $P$. The wire diameter $d=2r$ is the diameter of the fiber/wire. The mean diameter \textit{D} is the diameter of the helix $D = 2 R$. The pitch $p$ is the distance between the center of the rod in adjacent coils as shown in \cref{fig: Geometry}. The spring index \textit{SI} is given by the ratio between the mean coil diameter and the wire diameter $SI = D/d$.

The second set of parameters is mostly used in Kirchoff Rod Theory and also adopted in this work as shown in \cref{fig: Geometry}: radius of the coil $R$, normalized rod length per coil $l$, normalized pitch $p$ and internal twist $\tau_i$. Normalization of the rod length per coil and pitch is achieved by dividing these two parameters by $2\pi$. 

Finally, there are parameters mostly used in knot theory: linking number, twist number, and writhe. These parameters are often confused with the twist defined in the last paragraph The linking number $Lk$ corresponds to the amount of end rotation $N$ through which the external torque $L$ does work. For a twisted helix, it is given by $Lk=S(1/l+\tau_i)/2\pi=N$. The total twist number $Tw$ is the integral along the rod of the twist rate $\tau/2\pi$. For a twisted helix, it is given by $Tw=S(\sin{\alpha}/l+\tau_i)/2\pi$. Finally, writhe $Wr$ is a property of the space curve of the center line given by the difference between the twist number and linking number $Wr=Lk-Tw$. For a twisted helix, it is given by $Wr=\frac{S}{2\pi}(\frac{1-\sin{\alpha}}{l})$.

\newpage
\section{Fiber Differential Scanning Calorimetry (DSC) Test}
\label{appen: Fiber DSC}

    \begin{figure}[ht]
    \centering
    \includegraphics[width=1\textwidth]{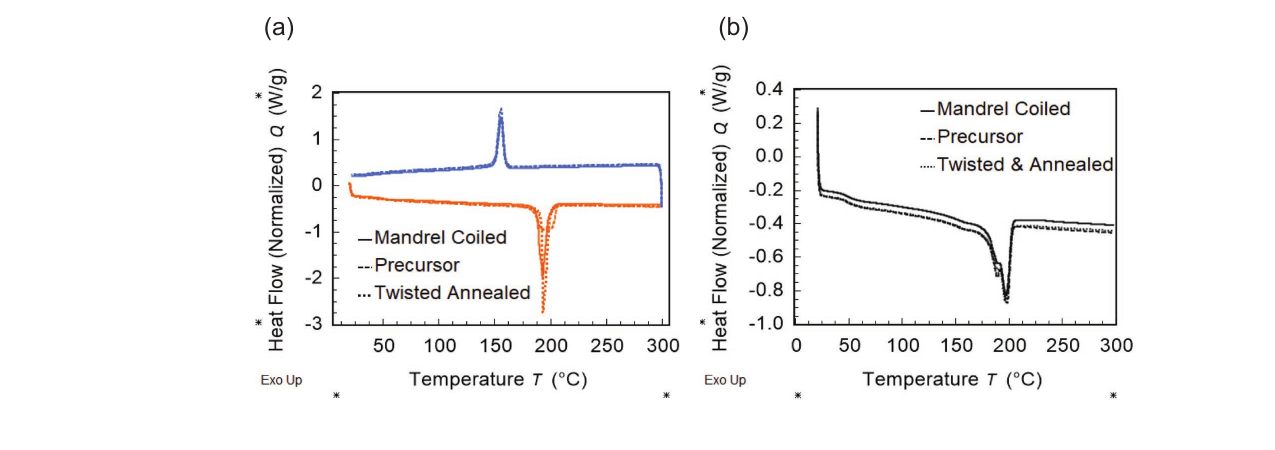}

    \caption{DSC curve of the precursor fiber, twisted and annealed straight fiber and mandrel coiled fiber. The test is conducted with 10 $^\circ C$/min from 25 $^\circ C$ to 300 $^\circ C$ and from 300 $^\circ C$ to 25 $^\circ C$. (a) DSC curve of the first heating and cooling cycle;
    (b) DSC curve of the second heating cycle.
    }
    \label{fig: DSCFig}
\end{figure}
We perform a DSC analysis of the fiber and analyze its thermal transitions such as glass transitions, crystallization and melting. The test is performed with a Discovery 2500 Differential Scanning Calorimeter. Three samples are being tested: precursor fiber, twisted and annealed straight fiber used in \cref{sec: Role of Micro} and mandrel coiled TCPA used in \cref{sec: Model validation}. All fiber samples are dried for 12 hours in a vacuum oven at room temperature. The samples are analyzed over the temperature range ambient to 300 °C with a ramping rate 10 °C/min for both heating and cooling with the atmosphere around the sample being nitrogen. Since we are specifically interested in the thermal history of the sample, we will focus on the first heating and cooling cycle as shown in \cref{fig: DSCFig}a. For the precursor fiber, the peak temperature of melting is 194 °C with melting endotherm of 55 J/g. The \% crystallinity based on 230 J/g is then 24 \%. For both twisted and annealed straight fiber and the mandrel coiled TCPA sample, the melting peak is at 193 °C and the \% crystallinity is 32 \%. The glass transition temperature of the samples is determined through the second heating cycle. As shown in \cref{fig: DSCFig}b, the glass transition temperature of the three samples is around 45 °C.

\newpage
\section{Water Isobaric Test}
\label{appen: Water isobaric}
\begin{figure}[ht]
    \centering
    \includegraphics[width=1\textwidth]{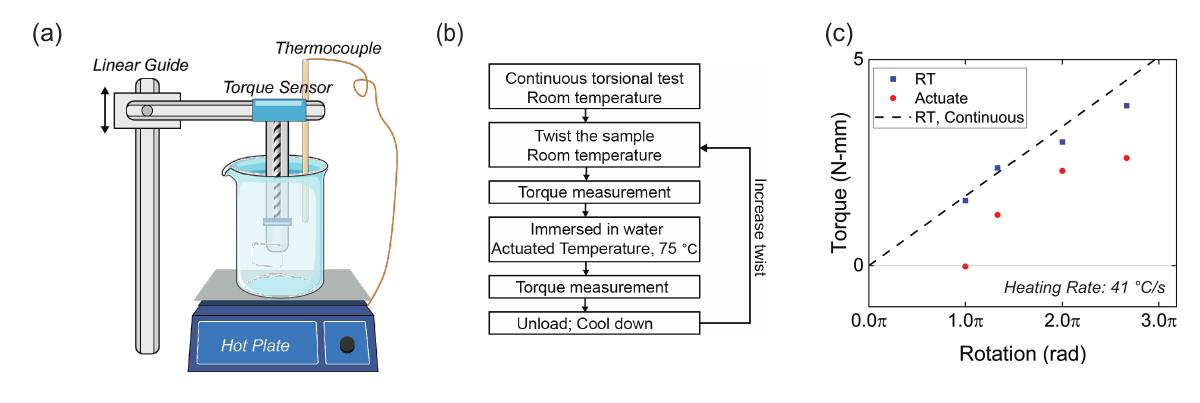}
    \caption{(a) The schematic of the experimental setup.  A beaker containing water was heated by a hot plate with the thermocouple, ensuring a constant temperature. The fiber sample was fixed to a vertical holder equipped with a torque sensor at the upper end. The fiber holder was then connected to a linear guide, enabling stable immersion; 
    (b) Experimental procedure;
    (c) The reaction torque of the sample before and after actuation was noted with markers and the room temperature torsional test in a dashed line.
     }
    \label{fig: WaterTestSetup}
\end{figure}

To investigate the thermomechanical properties of nylon fiber under fast heating, a specialized setup was designed, as depicted in \cref{fig: WaterTestSetup}a. A beaker containing water was heated by a hot plate with a thermocouple, ensuring a constant temperature. The fiber sample was fixed to a vertical holder equipped with a torque sensor at the upper end. The fiber holder was then connected to a linear guide, allowing for stable immersion. The tested sample was prepared by twisting and annealing the straight nylon fiber, following the procedure outlined in \cref{sec: Role of Micro}. The experimental procedure for the torsional test is described in \cref{fig: WaterTestSetup}b. Initially, the fiber was twisted by a certain degree at room temperature to measure its torsional reaction. Subsequently, with the fiber twisted, it was immersed in water using the linear guide to measure the torque at the actuated state. The fiber was then removed from the water and unloaded before the next test. The test results are plotted in \cref{fig: WaterTestSetup} with a dashed line showing the reaction torque of a continuous torsional test at room temperature. The test results indicate that the reaction torque of the fiber decreases upon immersion in water. However, it is worth noticing that after the fiber returned to room temperature, the torque might not fully recover to its initial pre-immersion value due to the relaxation of the fiber chains. With these tests, we are able to obtain the parameter of the system under a relatively fast heating rate, as shown in \cref{table: water}.

\newpage
\section{Optimization method}
We can obtain the control equation from \cref{eqn: CompletePotential} to \cref{eqn: CoupleEnergy2}. It's important to note that because Case A\&B are reduced forms of Case C, and Case 1 is a reduced form of Case 2, our focus should be on deriving the control equations for Case C2 and assigning a value of 0 to the parameters that are not in use for the other cases. By substituting \cref{eqn: TosionEnergyC} and \cref{eqn: CoupleEnergy2} into \cref{eqn: CompletePotential}, the total potential of the system can be written as following:

\begin{align*}
    &\Pi=W_K+W_R+U_B+U_T+U_C, \label{eqn: TotalPotentialappen}\\
    &W_K=-K(\tau_i+1/l-\tau_{i0}-1/l_0),\quad W_R=R (\sin{\alpha_0}-\sin{\alpha}),\\
    &U_B= \frac{1}{2} B(\kappa-\kappa_p)^2,\\
     &U_T= \frac{1}{2} (C(\tau-\tau_p)^2+C_{am} (\tau-\tau_0)^2) +C_3 \tau^3/3,\\
    &U_C=B_{C1}(\tau_i-\tau_{i1})(\kappa-\kappa_p)+B_{C2}(\tau_i-\tau_{i2})(\kappa-\kappa_p)^2.
\end{align*}

The equilibrium solution is then given by setting the partial differential of the total potential energy of the variable to be zero according to \cref{eqn: EquilEqnPartial}, given by following equations:

\begin{align}
    \begin{cases}
    K=C(\tau-\tau_p)+C_{am}(\tau-\tau_0)+C_3 \tau^2+B_{C1}(\kappa-\kappa_{p})+B_{C2}(\kappa-\kappa_p)^2\\
    R l\cos{\alpha}=(C(\tau-\tau_p)+C_{am}(\tau-\tau_0)+C_3\tau^2) \cos{\alpha}-(B(\kappa-\kappa_p)+B_{C1}(\tau_i-\tau_{i1}) +2 B_{C2}(\tau_i-\tau_{i2})(\kappa-\kappa_p) ) \sin{\alpha}\\
    K=(C(\tau-\tau_p)+C_{am}(\tau-\tau_0)+C_3\tau^2) \sin{\alpha}+(B(\kappa-\kappa_p)+B_{C1}(\tau_i-\tau_{i1}) +2 B_{C2}(\tau_i-\tau_{i2})(\kappa-\kappa_p)) \cos{\alpha}
    \end{cases}.
\end{align}

All material properties are determined through the torsional tests and listed in \cref{table: DMA} and \cref{table: water} except the coupling terms $B_{C1}$ and $B_{C2}$. The geometry parameters are measured manually after the sample is manufactured, which is also listed in  \cref{table: DMA} and \cref{table: water} except the internal twist $\tau_i$. This leads to either 2 (for Case 1) or 4 (for Case 2) unknow parameters parameters that need to be determined. In this work, they are optimized with Optuna \cite{Akiba2019Optuna:Framework} for each formulation by finding the minimum percent error. 5000 trials are run with TPE (Tree-structured Parzen Estimator) algorithm for the sampler and the Hyperband for the pruner. 

\newpage
\section{Tables of Parameters}
\label{appen: Simulation}

\begin{table}[ht]
\centering
\begin{tabular}{lllll}
\cline{1-3}
Total Linear Torsional Rigidity $C^0_{tot}$                  &5.5e-5  &    $Nm^2$\\
Second-Order Torsional Rigidity $C_3^0$        &1.65e-7  &    $Nm^2$\\
Crystallinity   $\Lambda$                          & 30 &    \%  \\
Thermal Coefficient of Linear Torsional Rigidity $\zeta_C$ &-7e-7  &   $Nm^{2\circ} C^{-1}$ \\
Thermal Coefficient of Second Order Torsional Rigidity $\zeta_{C3}$&-2.1e-9  &   $Nm^{2\circ} C^{-1}$ \\ 
Initial Internal Twist $\tau_{00}$&  50  &   $rad/m$ \\ 

\cline{1-3}
\end{tabular}
\caption{Simulation parameters - Torsional Fiber Test}
\label{table: Torsional Test}
\end{table}

\begin{table}[ht]
\centering
\begin{tabular}{lllll}
\cline{1-3}
Initial helix Angle                         & 0.27 & rad\\
Initial normalized fiber length per coil & 1.4 & mm\\
Temperature                         & 62 & $^\circ C$\\
Bending Rigidity                          &2.42e-4  &  $Nm^2$ \\
Total Torsional Rigidity                  &5.5e-5  &    $Nm^2$\\
Crystallinity                             & 30 &    \%  \\
Thermal Coefficient of Bending Rigidity   &  -3.1e-7& $Nm^{2\circ} C^{-1}$  \\
Thermal Coefficient of Torsional Rigidity &-7e-7  &   $Nm^{2\circ} C^{-1}$ \\ 
Thermal Coefficient of Second Order Torsional Rigidity &-2.1e-9  &   $Nm^{2\circ} C^{-1}$ \\
\cline{1-3} \\
\end{tabular}
\caption{Simulation parameters - DMA Test}
\label{table: DMA}
\end{table}

\begin{table}[ht]
\centering
\begin{tabular}{lllll}
\cline{1-3}
Initial helix Angle                         & 0.22 & rad\\
Initial normalized fiber length per coil & 1.2 & mm\\
Actuation Temperature &77& $^\circ C$ \\
Bending Rigidity                          &  1.4e-4&   $Nm^2$ \\
Total Torsional Rigidity                &  2.8e-5&  $Nm^2$    \\
Crystallinity                             &  30 & \%  \\
Thermal Coefficient of Bending Rigidity   & -2e-7 & $Nm^{2\circ} C^{-1}$   \\
Thermal Coefficient of Torsional Rigidity &  -8e-8& $Nm^{2\circ} C^{-1}$  \\ \cline{1-3}
\end{tabular}
\caption{Simulation parameters -Water Isobaric Test}
\label{table: water}
\end{table}

\end{document}